\documentclass[twocolumn,showpacs,preprintnumbers,amsmath,amssymb,superscriptaddress]{revtex4}
\usepackage{graphicx}% Include figure files
\usepackage{dcolumn}% Align table columns on decimal point
\usepackage{bm}% bold math
\usepackage[chatter]{rotating}
\usepackage{gensymb}

\usepackage{epstopdf}
\def\bea{\begin{eqnarray}}
\def\eea{\end{eqnarray}}

\newcommand{\dphi}{\ensuremath{\phi_\Delta}}

\newenvironment{color}[3]{
}{
}

\newcommand{\cyan}[1]      {\begin{color}{0}{1}{1}{#1}\end{colJor}}

\def\pp{\mbox{$p$-$p$}}

\def\auau{\mbox{Au-Au}}

\def\aa{\mbox{A-A}}
\def\ab{\mbox{A-B}}
\def\nn{\mbox{$N$-$N$}}

\def\pt{$p_t$}

\def\nch{$n_{ch}$}

\begin{document}

\setlength{\pdfpagewidth}{8.5in}
\setlength{\pdfpageheight}{11in}

% \setpagewiselinenumbers
% \modulolinenumbers[5]
% \linenumbers

\preprint{Version 2.3}

\title{Optimal modeling of 1D azimuth correlations in the context of
  Bayesian inference}

\author{Michiel B.~De Kock} %
\author{Hans C.~Eggers} %
\affiliation{Stellenbosch University and National Institute for
  Theoretical Physics (NITheP), ZA-7600 Stellenbosch, South Africa } %
\author{Thomas A. Trainor} %
\affiliation{CENPA 354290, University of Washington, Seattle,
  Washington 98195, United States}

%%%%%%%%%%%%%%%%%%%%%%%%%%%%%%%%%%%%%%%
\date{Version September 2015}

\begin{abstract}
  Analysis and interpretation of spectrum and correlation data from
  high-energy nuclear collisions is currently controversial because
  two opposing physics narratives derive contradictory implications
  from the same data---one narrative claiming collision dynamics is
  dominated by dijet production and projectile-nucleon fragmentation,
  the other claiming collision dynamics is dominated by a dense,
  flowing QCD medium.  %
  Opposing interpretations seem to be supported by alternative data
  models, and current model-comparison schemes are unable to
  distinguish between them.
  There is clearly need for a convincing new methodology to break the
  deadlock.
  In this study we introduce Bayesian Inference (BI) methods applied
  to angular correlation data as a basis to evaluate competing data
  models. For simplicity the data considered are projections of 2D
  angular correlations onto 1D azimuth from three centrality classes
  of 200 GeV \auau\ collisions. We consider several data models
  typical of current model choices, including Fourier series (FS) and
  a Gaussian plus various combinations of individual cosine
  components. We evaluate model performance with BI methods and with
  power-spectrum (PS) analysis.
  We find that the FS-only model is rejected in all cases by Bayesian
  analysis which always prefers a Gaussian. A cylindrical quadrupole
  $\cos(2\phi)$ is required in some cases but rejected for
  0-5\%-central \auau\ collisions.  Given a Gaussian centered at the
  azimuth origin ``higher harmonics'' $\cos(m\phi)$ for $m > 2$ are
  rejected.  A model consisting of Gaussian + dipole $\cos(\phi)$ +
  quadrupole $\cos(2\phi)$ provides good 1D data descriptions in all
  cases.
\end{abstract}


\maketitle

% % % % % % % % % % % % % % %
\section{Introduction}

A significant and persistent problem has emerged concerning models for
high-energy nucleus-nucleus (\aa) collision data from the relativistic
heavy ion collider (RHIC) and the large hadron collider (LHC).
Distinct classes of data models with divergent physics implications
are invoked to support two narratives: a high-energy physics
(HEP)/jets narrative in which the essential phenomenon is dijet
production~\cite{ua1,kll,sarc,hijing,kn} and a quark-gluon plasma
(QGP)/flow narrative in which the essential phenomenon is a flowing
dense QCD medium or QGP and dijets play no significant
role~\cite{perfliq1,perfliq2}.

The HEP/jets narrative emerges spontaneously from an analysis program
based on spectrum and correlation data models derived from the
observed differential structure of available
data~\cite{ppprd,hardspec,porter2,porter3,axialci,anomalous}.
In contrast, models emerging from the QGP/flow narrative tend to rely
on theoretical motivations coupled with data and information selection
(e.g.\ \pt\ cuts, preferred \aa\ centralities, ratio
measures)~\cite{poskvol,2004,blastwave,trigger,staras,starraa}.  A
comparison of RHIC results and interpretations is presented in
Ref.~\cite{review}.

For example, 2D angular correlations from high-energy nuclear
collisions include only a few structures common to all collisions from
\pp\ to central \auau\ at RHIC energies. A simple mathematical model
of those structures describes almost all data accurately with no
significant residual
structure~\cite{porter2,porter3,axialci,anomalous}.
No theoretical assumptions motivated the data model. Three of the four
principal model elements have been interpreted \textit{post facto} as
representing dijet production and projectile-nucleon
dissociation~\cite{jetspec,jetspecth,pptheory}. Interpretation of the
fourth element, an independent azimuth quadrupole, remains in
question~\cite{azimuth1,gluequad,quadspec,davehq,davehq2,nov2,nohydro}.
Differential analysis of hadron \pt\ spectra reveals two components
modeled by simple functions~\cite{ppprd,hardspec}. One component is
identified with fragments from dijets described quantitatively by QCD
calculations~\cite{fragevo}. Most spectrum and correlation structures
appear to be consistent with the HEP/jets narrative.
Alternative models motivated by the QGP/flow narrative include
quantity $v_2$ [Fourier coefficient of function $\cos(2\phi)$ fitted
to 1D projections of 2D angular correlations] interpreted to represent
elliptic flow~\cite{poskvol,2004}, a blast-wave spectrum model
interpreted to measure radial flow~\cite{blastwave}, spectrum ratio
$R_{AA}$ interpreted to indicate jet quenching within a dense QCD
medium~\cite{starraa}, and dihadron correlation analysis via
background subtraction interpreted to represent jet
structure~\cite{staras,trigger,tzyam}. ``Higher harmonic'' flows have
been inferred recently from azimuth distributions via Fourier-series
models~\cite{gunther,luzum,lhcharm}.

The same underlying particle data are therefore characterized and
interpreted with competing mathematical models applied to different
data selections, variables and measured quantities. Judgments on the
validity and relative merits of competing data models have relied
historically on comparisons of minimum-$\chi^2$ values and qualitative
arguments based on consistency of a given narrative across selected
measured quantities. While such an approach might suffice when the
underlying physical processes and models are simple, the complexity of
\aa\ phenomenology and lack of consistent quantitative criteria have
impeded progress in resolving conflicts.

To address this problem we require a formal context in which competing
data models are evaluated on a statistically sound basis, and a
``best'' model may be selected that either does not rely on unspoken
{\em a priori} physics assumptions or renders such assumptions
quantifiable.
We suggest that this context exists in the form of Bayesian Inference
(BI) which provides both a formal mathematical framework and the
necessary concepts to represent prior knowledge, evaluate candidate
models for different parameter values and thereby establish value
judgments on models as a
whole~\cite{bayes1,bayes2,bayes3,mackay}. Each data model is rated not
only by how well it describes some data or how much data it describes
well, but also by the ``cost'' of the model in terms of complexity and
parameter number (Occam penalty) and associated physical assumptions.

In this study we focus on 1D projections onto azimuth $\phi$ of 2D
angular correlations reported in Ref.~\cite{anomalous}, currently one
of the most contentious areas of RHIC/LHC data analysis.  We consider
several popular data models and evaluate them according to BI methods
to determine whether a uniquely preferred data model can be
established without recourse to {\em a priori} physics assumptions.

This article is arranged as follows: Section~\ref{bayes1} presents the
basics of Bayesian Inference.  Section~\ref{fourier} describes Fourier
power spectra (PS) and their properties.  Section~\ref{meths}
summarizes analysis methods applied to correlation data.
Section~\ref{corr} introduces the correlation data used for this
study.  Sections~\ref{bin10a}, \ref{bin8a} and \ref{bin0} apply BI and
PS methods to azimuth projections from three centralities of 200 GeV
\auau\ collisions.  Section~\ref{system} presents
systematic-uncertainty estimates.  Sections~\ref{disc} and~\ref{summ}
present discussion and summary.  Appendices A and B consider the
geometry of BI analysis and periodic peak arrays respectively.

\section{Bayesian inference} \label{bayes1}

Bayesian Inference addresses the problem of relating parametrized
model functions to available data in an optimal manner. Given specific
data values the best set of parameter values for each model is
determined based on the {\em likelihood} function.
Several models are then compared based on each model's
\textit{evidence}, an integral measure defined below. The most
plausible and therefore preferred data model produces the largest
evidence value. %

\subsection{The $\bf \chi^2$ measure and model fits to data}

In the present study we focus on aspects of Bayesian Inference that
correspond directly with the methodology of $\chi^2$
minimization. Given a set of $N$ data points $(x_n,y_n)$ with
experimentally determined standard errors $\sigma_n$ on $y_n$ the
conventional $\chi^2$ statistic evaluating the goodness of fit of
model function $f(x\,|\,w)$ with $K$ parameters $w_k$ is
\begin{equation} 
  \chi^2 = \sum_{n=1}^N \left( \frac{y_n - f(x_n\,|\,w)}{\sigma_n}\right)^2 .
\end{equation} 
As stated in Ref.~\cite{dekock14} the $\chi^2$ measure assumes a
Gaussian distribution of data-sample fluctuations about mean values
which we accept as a reasonable approximation for 1D RHIC/LHC data
projections. In what follows model functions are represented by
$D(w)$, a vector function mapping parameter space $w$ to data space
$D$.

Most model comparisons are based on $\chi^2$/DoF, where the number of
{\em fit} degrees of freedom (DoF) is assumed to be the number of data
points $N$ minus the number of free model parameters $K$. Minimizing
$\chi^2$ without considering the fit DoF is clearly misleading since
there are infinitely many models with $K {=} N$ free parameters that
might describe the same $N$ data points with $\chi^2= 0$
\cite{jeffreys}.
We require a mechanism to penalize excess model parameters such that a
simple few-parameter model that describes the data well may be favored
over more-complex models. That mechanism exists in the form of
Bayesian Inference.

\subsection{Logical and rational inference}

Distinction may be drawn between logical inference on the one hand, in
which nominally-valid conclusions are drawn via a logical chain of
argument from premises assumed to be true and rational inference on
the other, in which patterns or events (i.e.\ data) are used to
improve our understanding of the physical system, either augmenting or
displacing previous understanding. %
Both the acquired data and the modified understanding may be uncertain
to some degree as measured by probabilities. Rational inference
includes induction, in which newly-acquired data are employed to
formulate or refine a model, and deduction in which a fixed model is
used to predict values of data not yet acquired~\cite{inference}.

Bayesian Inference is a formal recipe for rational inference based on
Bayes' theorem~\cite{jeffreys,jaynes}.  ``Understanding'' in this
context means that reality in the form of data or data-derived
quantities is well described by a parametrized model.
A given set of parameter values predicts a specific set of {\em
  possible} data values.  Previous understanding including
uncertainties is represented by the {\em prior}, a probability
distribution function (PDF) on possible parameter values.  As new data
are acquired BI provides a means to update the PDF on model parameters
to effect improved understanding in the form of the posterior PDF,
thereby refining the model by reducing the volume of its parameter
space or falsifying the model altogether if the new data fall outside
the model's predicted data volume.

\subsection{The probability chain rule and Bayes' theorem} \label{chain}

Bayesian Inference is based on relations among joint, conditional and
marginal PDFs and related unnormalized functions distributed on data
and model-parameter spaces~\cite{kendall}. External factors common to
all models that may influence the inference process are represented by
a comprehensive parameter set $Q$ suppressed below.  Our notation
follows that in Refs.~\cite{mackay} and \cite{jaynes}.

A model $H$ is defined by a joint PDF $p(wD|HQ) \rightarrow p(wD|H)$,
where $w$ and $D$ are multidimensional spaces representing
model-parameter values and data values. The corresponding conditional
PDFs are $p(w|DH)$ and $p(D|wH)$, and the marginal PDFs are $p(w|H)$
and $p(D|H)$. The probability chain rule provides factorizations in
the form $p(wD|H) = p(w|DH)p(D|H) = p(D|wH)p(w|H)$.
Bayes' theorem (BT) can then be expressed in either of two forms
\begin{eqnarray} \label{bayes}
  p(w|DH) &=& \frac{p(D|wH)p(w|H)}{p(D|H)}
  \\ \nonumber
  p(D|wH) &=& \frac{p(w|DH)p(D|H)}{p(w|H)},
\end{eqnarray}
both of which are valid descriptions of a joint PDF. However, only the
first line is applicable to BI analysis that proceeds from specific
data values to improved parametrized data model, a unique BT
application.

\subsection{Prior and posterior PDFs -- model fits}

As applied to BI analysis some quantities in the first line of
Eq.~(\ref{bayes}) must be defined more specifically. In this
application quantity $D$ is not a variable on the space of all
possible data; it is a specific set of data values $D^*$ with
uncertainties or errors $\sigma_D$.
Factor $p(D|wH)$, a normalized conditional PDF on data space $D$, is
redefined as the {\em likelihood} function $L(D^*|wH)$ on parameter
space $w$ for model $H$ given specific data $D^*$ and model function
$D(w)$.  $p(w|H)$ is the {\em prior} PDF on model parameters $w$
determined before data $D^*$ are available.  $p(w|D^*H)$ is the {\em
  posterior} PDF on parameters $w$ given the new data.  Denominator
$p(D|H)$, also a PDF on space $D$, is redefined as the {\em evidence}
(a number) for model $H$ given specific data $D^*$ which we denote by
the symbol $E(D^*|H)$.  With those more-specific definitions the
version of Bayes' Theorem used for BI is
\begin{eqnarray} \label{inference}
p(w|D^*H) &=& \frac{L(D^*|wH)}{E(D^*|H)}p(w|H),
\end{eqnarray}
which can be read as ``A posterior PDF on $w$ is derived from a prior
PDF given data $D^*$, likelihood L and evidence E.'' Any change
between prior and posterior represents {\em information} acquired by
the model from the data. The result is an updated PDF on model
parameters determined by newly-acquired specific data values
$D^*$. The posterior PDF on parameters $w$ provides considerably more
information about the model than the best-fit parameter set $\tilde w$
and uncertainties $\sigma_w$ derived from conventional $\chi^2$ model
fits to data.

\subsection{Model comparisons and evidence} \label{modcompare}

Beyond determining posterior PDFs on parameters $w$ Bayes' Theorem can
be used on a higher level for comparisons among competing data models
in the form
\begin{eqnarray} \label{phd} p(H|D^*) &=& \frac{E(D^*|H) p
    (H)}{p(D^*)},
\end{eqnarray}
where $p(H|D^*)$ is the {\em plausibility} of model $H$ given data
values $D^*$ and $p(H)$ is the prior model probability within some
assumed context represented by $Q$ (suppressed).
The main goal of this study is comparison of competing model functions
$D(w)$ with all other BI elements maintained as similar as possible.

Evidence $E$ is just a normalization parameter in
Eq.~(\ref{inference}), but its absolute numerical value is important
for model comparisons. Because the likelihood is usually a peaked
function on $w$ with single mode near some optimal parameter values
$\tilde w$ the evidence defined in the first line below can be
represented by Laplace's approximation in the second
line~\cite{laplace}
\begin{eqnarray} \label{evid} 
  E(D^*|H) &=& \int dw L(D^*|wH) p(w|H) \\
  \nonumber 
  &\approx& L(D^*|\tilde w H) \sqrt{(2\pi)^K \det C_K}\,
  p(\tilde w|H),
\end{eqnarray}
where $L(D^*|\tilde w H)$ is the maximum likelihood and
$C_K(D^*|\tilde w H)$ is the covariance matrix for model function
$D(w)$ with $K$ parameters. The negative log evidence is \
\begin{eqnarray} \label{le}
-2LE &\approx& \chi^2(D^*|\tilde w H)  + 2I(D^*|H) + \text{constant}
\end{eqnarray}
with usual $\chi^2$ parameter, and information $I$ is defined by
\begin{eqnarray} \label{infoeq1}
I(D^*|H) &=& -\ln\left[ \sqrt{(2\pi)^K \det C_K}~ p(\tilde w|H) \right],
\end{eqnarray}
the information gained by model $H$ from specific data
$D^*$. Information is the log of a volume ratio as discussed in the
next subsection. In general $\chi^2$ decreases and $I$ increases as
parameter-number index $K$ increases.  The sum $-2LE$ should then have
a minimum corresponding to the maximum evidence for a specific
model. For an optimized {\em predictive} model (e.g.\ a theory)
$I\approx 0$ and $\chi^2 \approx$ fit DoF (= data DoF $N $ minus model
DoF $K$).

Quick and easy comparisons between two models $H_1$ and $H_2$ can be
obtained by calculating the evidence ratio $E(D^*|H_1)/E(D^*|H_2)$,
also known as an odds ratio. Assuming equal model priors $p(H_1) =
p(H_2)$ the {\em Bayes Factor} is \cite{bayes3, dekock11}
\begin{eqnarray}
\label{bafact}
B_{12} &=& \ln\frac{p(H_1|D^*)}{p(H_2|D^*)} = \ln\frac{E(D^*|H_1)}{E(D^*|H_2)}.
\end{eqnarray}
Comparisons among more than two models indexed by $l$ are effected by
\begin{eqnarray}
p(H_l|D^*) &=& \frac{E(D^*|H_l) p(H_l)}{\sum_l E(D^*|H_l) p(H_l)},
\end{eqnarray}
where $\sum_l E(D^*|H_l) p(H_l)$ replaces $p(D^*)$ in
Eq.~(\ref{phd}). The model priors $p(H_l)$ could be set equal assuming
ignorance, but in practice assigned model priors may differ sharply
among competing models, possibly reflecting strong prejudices.

Our use of differences in log Evidence (Bayes factors) rather than
isolated values is consistent with the use of Likelihood ratios (e.g.\
Neyman-Pearson approach). Evidence ratios are an improvement on
Likelihood ratios because the latter assume delta-function priors.

\subsection{Bayesian priors and Information} \label{occam}

Information is generally defined as the logarithm of a volume ratio,
the volumes being subsets of some space of alternatives before and
after a message (data) is received conveying information. For
instance, if a message reduces the number of possible alternatives by
factor 2 then the amount of information received is $\log_2(V_1/V_2 =
2) = 1$: one ``bit'' of information is provided by the message.
Several definitions of information have been formulated (e.g.\
Shannon, R\'enyi), and the precise correspondence to a volume ratio
varies from case to case. In some cases the terms ``information'' and
``entropy'' may be used interchangeably such that for example
``information gain'' may represent the difference between two
entropies.

In Eq.~(\ref{infoeq1}) factor $p(\tilde w|H)$ is related to the prior
volume $V_w(H)$ of a model parameter space and $\sqrt{(2\pi)^K \det
  C_K}$ approximates the posterior volume $V_w(D^*|H)$. Thus,
information $I(D^*|H)$ is defined here as the natural log of the prior
volume over the posterior volume. A prior PDF based on ignorance
(uniform or translation-invariant probability within some assumed
boundaries for each parameter) is estimated by the product
\begin{eqnarray} \label{priordelt}
p(w|H) \approx \prod_{k=1}^K \frac{1}{\Delta_{k}} \equiv \frac{1}{V_w(H)},
\end{eqnarray}
where the estimated $\Delta_{k}$ for amplitude parameters may be based
on differences of data extreme values, but the prior for angle
parameters depends on circumstances. In this study the condition
$\sigma_{\phi_\Delta} \in [0,\pi/2]$ is based on the definition of the
{\em same-side} peak at the azimuth origin.

Since typical correlation-structure amplitudes (e.g.\ peak-to-peak
excursions) are generally $< O(1)$ and given the assumed constraint on
the Gaussian width we assign $\Delta_k = 1$ for those cases. Given
certain algebraic relations it is reasonable to assume that cosine
coefficients and uncertainties may be substantially smaller on average
than the Gaussian amplitude and width. For all cosine components in
any model we assign $\Delta_{k} = 1/3$. Given those assignments the
{\em basic Model} (defined below) is somewhat disadvantaged (smaller
prior probability) compared to models based only on cosine
terms. Further discussion of prior construction is found in
Ref.~\cite{dekock14}.

The posterior volume is obtained from the determinant of the
covariance matrix $\det C_K$ which, in the absence of significant
covariances, is the product of the variances for the several model
parameters. Its square root is then the product of r.m.s.\ widths on
parameters, the posterior volume. In this study the Hessian (matrix of
second-order derivatives at maximum of the Likelihood function derived
from data $D^*$) is obtained, and the covariance matrix is constructed
from the Hessian elements.

The information defined in Eq.~(\ref{infoeq1}) permits a quantitative
expression of Occam's razor in two ways: (a) For a model with a large
prior volume in parameter space (representing many ``causes'', some
possibly unnecessary) a substantial reduction in the parameter volume
on encountering data $D^*$ automatically incurs an Occam penalty by
means of larger $I$. (b) The $K$-dependence of $I$ implies that while
models with more parameters may have a smaller $\chi^2$ and larger
likelihood, the extra parameters are also penalized by increased $I$
resulting in reduced overall model plausibility.

\section{Fourier power spectrum} \label{fourier}

The Fourier power spectrum (PS) is an alternative information measure
well understood in the context of signal processing. Comparison of PS
results with BI analysis may better convey the technical details and
interpretations of the latter.

The Wiener-Khinchin theorem~\cite{wk} states that the Fourier
transform of a two-particle autocorrelation is the corresponding power
spectrum of an underlying single-particle distribution. Data
autocorrelations $A(\phi_n)$ with $N$ elements are periodic,
symmetrized about 0 and $\pi$ and described by a PS with $m \in
[0,N-1]$ and $P_m = P_{N-m}$. The PS expansion of autocorrelation data
\begin{eqnarray} \label{wiener1}
A(\phi_n) &=&  \sum_{m=0}^{N-1} P_m \cos(m\phi_n)
\\ \nonumber
&=& P_0 + \sum_{m=1}^{N/2} P_m 2\cos(m\phi_n)
\end{eqnarray}
might be viewed as a model function from which the power-spectrum
elements $P_m$ could be determined by model fitting.  However, in this
study the PS elements are obtained directly by integrating the data
\begin{eqnarray} \label{coef}
P_m &\equiv& \frac{1}{2\pi} \int_{0}^{2\pi} \hspace{-.05in} d\phi_\Delta \cos(m \phi_\Delta) A(\phi_\Delta)
\\ \nonumber
&\rightarrow& \frac{1}{N} \sum_{n'=0}^{N-1}  \cos(m\phi_{n'}) A(\phi_{n'}).
\end{eqnarray}
Note that $A(0)$ is the ``total power'' $\sum_{m=0}^{N-1} P_m$ (with
$N/2 + 1$ independent elements), and $P_0$ is the mean value of the 1D
autocorrelation (which, for data histograms introduced below and used
in this study, is set to zero).

The power spectrum for a sample sequence may contain a deterministic
``signal'' component and a random (white) noise component. The signal
may be localized at smaller wave number (index $m$), while an
approximately flat white-noise spectrum is revealed at larger index
values if the sample rate or bin number (resolution) is large enough
(see Nyquist frequency limit below). The white-noise amplitude should
correspond to the estimated statistical (Poisson) error used in
$\chi^2$ fits to a sample sequence and to the r.m.s.\ error inferred
from fit residuals.

The Nyquist limit applied to periodic azimuth implies that the power
spectrum must be symmetric about the bin on $m$ containing $N/2$. For
$N=24$ there are then $12 + 1$ independent PS elements (including $m =
0$) and 13 unique autocorrelation data bins whose contents may be
correlated by one or more parent processes.
For the broader correlation structures considered here the bin number,
and therefore the Nyquist limit, is adequate. For the narrower
BE/electron peak (defined below) the bin number (hence angle
resolution) is insufficient, but that structure is not important for
this analysis.

Power spectra PS should be distinguished from Fourier series (FS-only)
fit models. A PS consisting of elements $P_m$ evaluated for all index
values $m \in [0,N/2]$ completely characterizes a data
autocorrelation. FS-only models have a varying number of elements
indexed by $k \leq K$, $K \in [1,N/2]$ being the number of parameters
for a model.

%%%%%%%%%%%%%%%%%%%%%%%%%%%%%
\section{Analysis methods} \label{meths}

High-energy nuclear collisions at the RHIC and LHC produce hadrons in
each collision ranging in number from a few to thousands (depending on
collision centrality) via several physical mechanisms. By studying
properties of hadron yields, spectra and correlations we seek to
identify and characterize the various underlying mechanisms. In this
study we apply BI methods to evaluate several mathematical models of
2D angular correlations projected to 1D azimuth. In this section we
summarizes basic analysis methods that produce the angular correlation
data and our strategy for BI evaluation of the data models.

\subsection{Kinematic variables and spaces} \label{kine}
 
High-energy nuclear collisions are described efficiently within a
cylindrical coordinate system $(p_t,\eta,\phi)$ where (relative to the
collision axis) $p_t$ is the transverse momentum, $\phi$ is the
azimuth angle from a reference direction and pseudorapidity $\eta = -
\ln [\tan(\theta/2)] \approx \cos(\theta)$ is a measure of polar angle
$\theta$, the approximation being valid near $\eta = 0$ ($\theta =
\pi/2$). A bounded detector angular acceptance is denoted by intervals
($\Delta \eta,\Delta \phi$) on the primary single-particle space
$(\eta,\phi)$.

In general, two-particle correlations are measured on the 6D space
$(p_{t1},\eta_1,\phi_1,p_{t2},\eta_2,\phi_2)$. \pt-integral angular
correlations are measured on the 4D space
$(\eta_1,\phi_1,\eta_2,\phi_2)$. Within a limited $\eta$ acceptance
and over $2\pi$ azimuth the angular correlation structure may be
approximately invariant along a sum axis $x_\Sigma = x_1 + x_2$
(stationarity). In that case averages along $x_\Sigma$ for each value
of the corresponding difference variable $x_\Delta = x_1 - x_2$
comprise an {\em autocorrelation} $A(x_\Delta)$. Angular correlations
on $(\eta,\phi)$ are then measured as 2D densities
$A(\eta_\Delta,\phi_\Delta)$ without significant loss of
information~\cite{inverse}.

\subsection{\aa\ centrality measures}

\aa\ collision centrality is measured by comparing a measured
minimum-bias (MB) event distribution on charge multiplicity $n_{ch}$
within some fiducial angular acceptance with a Glauber Monte Carlo
model of \aa\ collisions producing MB distributions on nucleon
participant number $N_{part}$ and \nn\ binary-collision number
$N_{bin}$~\cite{powerlaw}. The intermediary is the \aa\ fractional
cross section $\sigma / \sigma_0$. For the data employed in this study
centrality is designated by fractional cross section in percent, where
100\% refers to extreme peripheral collisions and 0\% refers to
head-on collisions. For the data employed in this study collision
events were sorted into eleven centrality bins: ten equal 10\%
centrality bins with the most-central 10\% bin split into two 5\%
bins.
The bins are numbered 0 (most peripheral) through 10 (most
central). The three (corrected) centrality intervals used in this
study are 0-5\% (bin 10), 9-18\% (bin 8) and 83-94\% (bin 0).

\subsection{Correlation measures}

Correlation structure is identified by comparing a 2D pair density
$\rho(\eta_\Delta,\phi_\Delta)$ with a reference density $\rho_{ref}$
representing no significant correlations or some uninteresting
background structure. $\rho_{ref}$ can be based for instance on a
factorization assumption ($\rho_{ref} = \rho_0^2$) or a distribution
of mixed pairs formed from different but similar sample events
($\rho_{ref} = \rho_{mix}$). The difference $\Delta \rho = \rho -
\rho_{ref}$ should reveal correlation structure of interest.

Correlation structure may have several components arising from
different collision mechanisms. Correlation amplitudes may vary with
collision conditions in characteristic ways, for instance proportional
to $n_{ch}$, $N_{part}$, $N_{bin}$ or some combination. As a
placeholder we define a {\em per particle} measure $\Delta \rho /
\sqrt{\rho_{ref}}$ since $\rho_{ref} \approx \rho_0^2$ according to a
factorization assumption, and $\rho_0 = d^2n_{ch} /d\eta \dphi$ is the
mean single-particle charge density near the angular
origin. Practically speaking the correlation measure is obtained as
\begin{eqnarray}
  \frac{\Delta \rho}{\sqrt{\rho_{ref}}} &\equiv& \rho_0 \left[\frac{\rho}{\rho_{mix}} - 1 \right]
\end{eqnarray}
where the ratio inside the square brackets reduces certain
instrumental effects~\cite{anomalous}. In what follows we refer to
symbol $A$ to simplify notation.

A 2D autocorrelation in the form $\Delta \rho / \sqrt{\rho_{ref}}
\rightarrow A$ is a density defined with the prefactor $d^2n_{ch}/
d\eta d\phi$. When integrated over $\eta_\Delta$ the autocorrelation
is a density on $\phi_\Delta$ defined by prefactor
$dn_{ch}/d\phi$. Integration of $A(\phi_\Delta)$ over the azimuth
acceptance should then give $2\pi \bar A \approx 0$ since $\rho_{mix}$
has the same pair number as $\rho$ by construction.

\subsection{Bayesian Inference strategy}

For each 1D data histogram we construct a PS as a reference for BI
analysis and identify within the PS the signal and noise
components. PS structure can be related 1-to-1 with BI elements,
helping to clarify interpretation of the latter. Based on results from
Ref.~\cite{anomalous} we compare the PS for a fitted 1D Gaussian with
each data PS.

For each model function $D(w)$ we obtain the minimum $\chi^2$ (maximum
likelihood describing fit quality) and information $I$ (derived from
priors and covariance matrix) from fits to data histograms. We obtain
evidence $E$ for each model from a combination of minimum $\chi^2$ and
information $I$.  Competition between $\chi^2$ and $I$ contrasts
goodness of fit (via $\chi^2$) with quantitative assessment of
model-parameter ``cost'' or {\em Occam penalty} (via $I$). One model
function may achieve a quantitatively better fit to data than another
model, but at the cost of extra model parameters that may favor the
second model overall.

We emphasize that the number of data DoF in this study is small, only
11 for the projected 1D histograms analyzed here compared to the
original 2D histograms with 169 DoF. The small number of data DoF
presents unique challenges for data modeling and BI evaluation.

%%%%%%%%%%%%%%%%
\section{Correlation data and models} \label{corr}

The data we consider were published in the form of 2D binned
histograms (autocorrelations) derived from 1.2M 200 GeV \auau\
collision events sorted into eleven centrality classes based on
charged-particle multiplicity \nch~\cite{anomalous}. Depending on
centrality each collision event may include from a few to more than a
thousand charged particles within the detector acceptance $(\Delta
\eta,\Delta \phi) = (2,2\pi)$.

In the present study we consider 1D projections of the 2D histograms
onto azimuth difference $\phi_\Delta$ represented as $\phi$ to
simplify notation. The histogram bin size on azimuth is $\delta \phi =
2\pi / N$ ($N =$ 24 bins). The position variable is then $\phi_n = n\,
2\pi / N$ with $n \in [0,N-1]$. The conjugate index for a PS
(Sec.~\ref{fourier}) is $m \in [0,N-1]$. The argument of PS cosines is
$m \phi_n = mn\, 2 \pi /N$.
The bin size has been optimized to match the observed correlation
structure and provides sufficient resolution to retain all information
in the data, as indicated for instance by the power spectrum in
Fig.~\ref{power1}.

The 2D data are symmetrized on both $\eta_\Delta$ and
$\phi_\Delta$. Thus, only one quadrant of each 2D histogram is
unique. The statistical errors on $\phi_\Delta$ are uniform except for
bins at 0 and $\pi$ where they are $\sqrt{2}$ larger. The errors on
$\eta_\Delta$ are strongly varying due to the triangular pair
acceptance on $\eta_\Delta$, with the largest errors at the acceptance
edges $|\eta_\Delta| \approx 2$. As noted, the 2D correlation
histograms sum to zero by construction.  We also adjust the 1D
projections onto $\phi_\Delta$ to zero sum leading to one less data
DoF (12).

\subsection{Correlation data histograms}

\begin{figure}[h]
  \includegraphics[width=1.65in,height=1.58in]{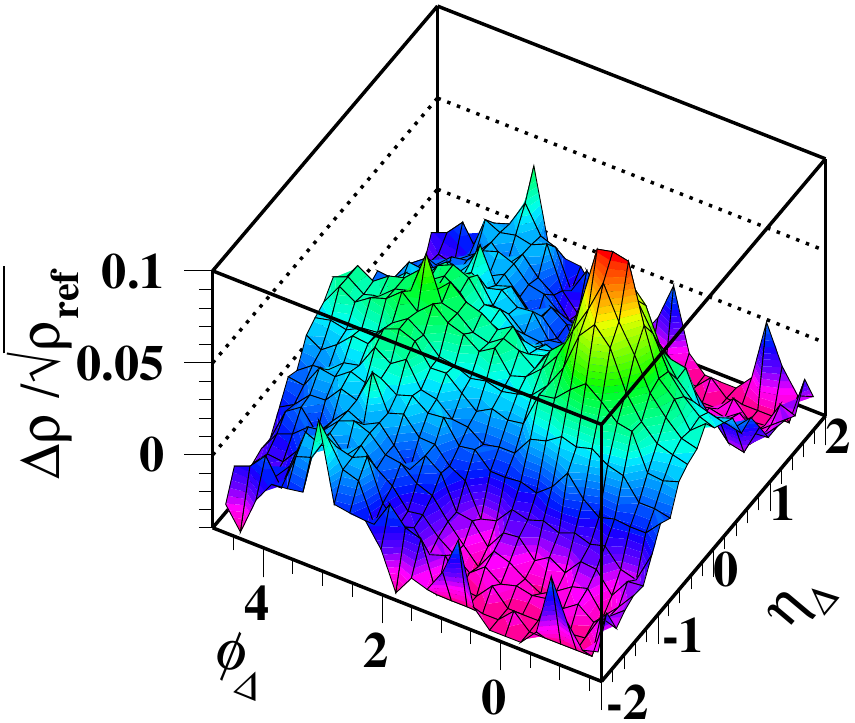}
  \put(-90,99) {\bf (a)}
 \includegraphics[width=1.65in,height=1.58in]{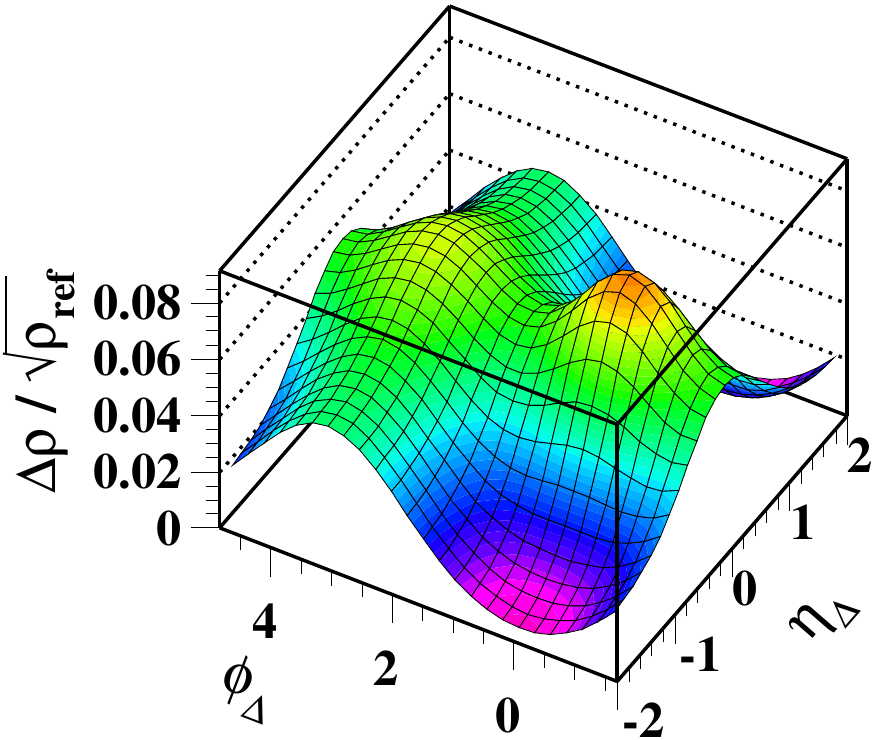}
 \put(-90,99) {\bf (b)} \\
  \includegraphics[width=1.65in,height=1.58in]{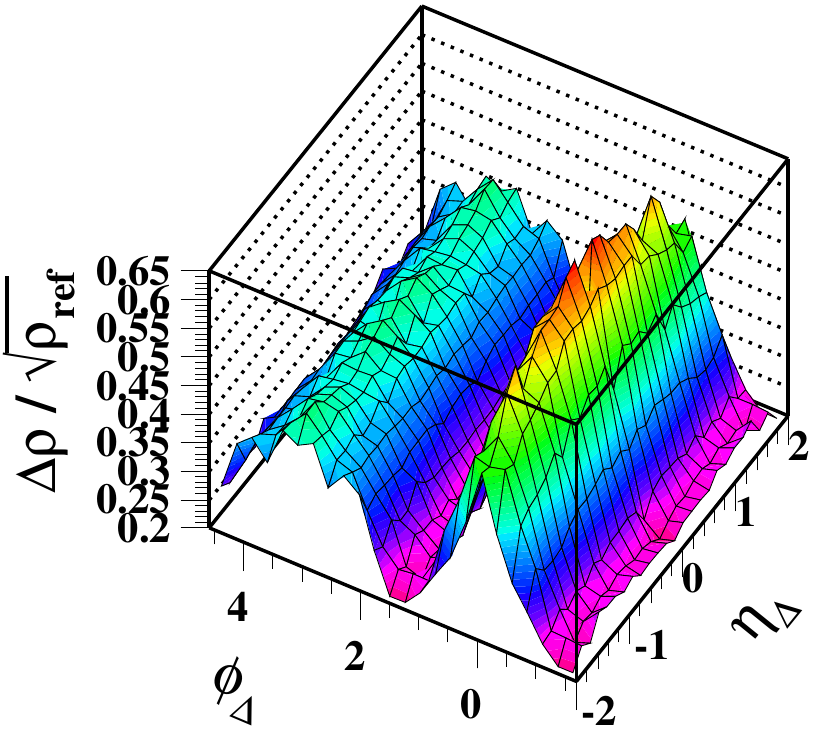}
  \put(-90,99) {\bf (c)}
  \includegraphics[width=1.65in,height=1.58in]{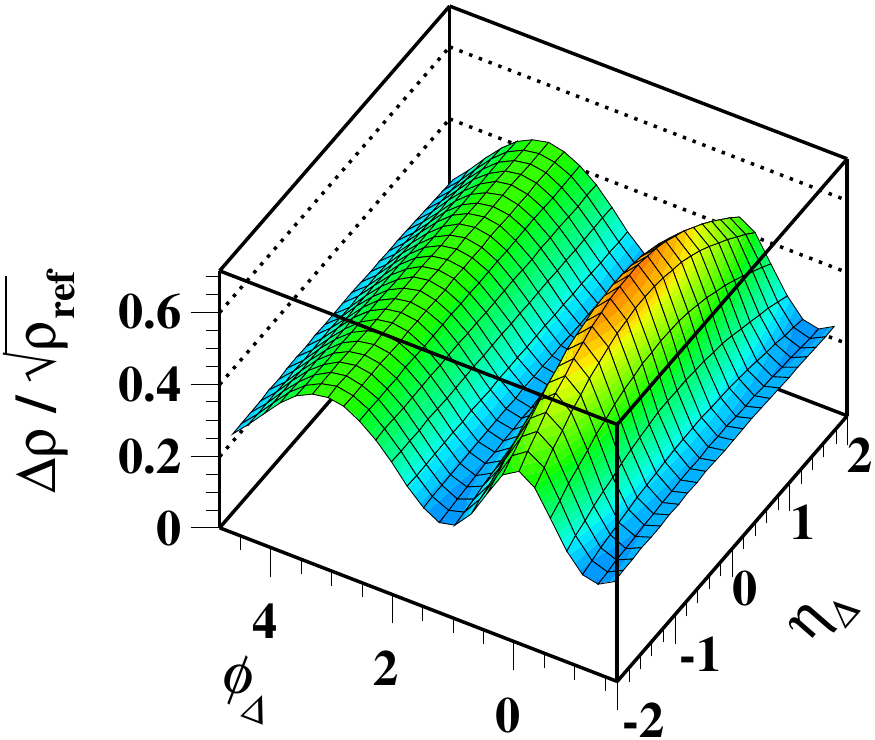}
  \put(-90,99) {\bf (d)}
  \caption{\label{fig1} (Color online) Left: 2D angular
    autocorrelations from 200 GeV \auau\ collisions for (a) 83-94\%
    ($\sim$\nn\ collisions) and (c) 0-5\% centralities.  Right:
    Two-dimensional model fits to the histograms in the left panels
    obtained with Eq.~(\ref{2dmodel}).  } %flow23-300-xx, ppcms50a-x
\end{figure}

Figure~\ref{fig1} (left panels) shows 200 GeV \auau\ 2D angular
correlations for centrality bin 0 (83-94\%, $\approx$ \nn\ collisions)
and bin 10 (0-5\%).
Within the STAR TPC acceptance the \pt-integral correlation data from
\auau\ collisions include four principal components: (a) a same-side
(SS) 2D peak at the origin on $(\eta_\Delta,\phi_\Delta)$ well
approximated by a 2D Gaussian for all \pt-integral data, (b) an
away-side (AS) 1D peak on azimuth well approximated by an AS dipole
$[\cos(\phi_\Delta - \pi) + 1]/2$ for all data and uniform to a few
percent on $\eta_\Delta$ (having negligible curvature), (c) an azimuth
quadrupole $\cos(2\phi_\Delta)$ also uniform on $\eta_\Delta$ to a few
percent over the full angular acceptance of the STAR TPC, and (d) a
narrow 1D peak on $\eta_\Delta$. There is also a sharp 2D exponential
peak at (0,0). That phenomenological description does not rely on
physical interpretations of the components.

%%%%%%%%%%
%%%%%%%%%%

Based on subsequent comparisons of observed data systematics with
theory the components (a) and (b) together are interpreted to
represent minimum-bias dijets~\cite{fragevo,anomalous}. Component (c)
has been conventionally attributed to elliptic
flow~\cite{2004}. Component (d) is attributed to projectile-nucleon
dissociation. And the 2D exponential is attributed to Bose-Einstein
(quantum) correlations and charge-neutral electron pairs from
photoconversions (denoted as the BE/electron peak).

\subsection{Correlation data models} \label{models}

The fit methods employed here are based on the the non-Fisherian
ansatz that data can be represented as the sum of a hypothesis (any
competing data parametrization) plus noise.
2D histograms from Ref.~\cite{anomalous} [e.g.\ Fig.~\ref{fig1} (a)
and (c)] were fitted with a data model including several elements
applicable to higher RHIC energies and all \auau\ centralities. The
11-parameter model is
\begin{eqnarray} \label{2dmodel} A(\eta_\Delta,\phi_\Delta) & = &
  A_{2D} \exp\left\{- \frac{1}{2} \left[ \left( \frac{\phi_{\Delta}}{
          \sigma_{\phi_{\Delta}}} \right)^2 \hspace{-.05in} + \left(
        \frac{\eta_{\Delta}}{ \sigma_{\eta_{\Delta}}} \right)^2
    \right] \right\} \\ \nonumber &+& A_{\rm D} [\cos(\phi_\Delta -
  \pi) + 1]/2 + A_0 \\ \nonumber &+& A_{\rm Q} 2\cos(2\, \phi_\Delta)
  \hspace{-.03in} + \hspace{-.03in} A_\text{soft}\,
  \exp\left\{-\frac{1}{2} \left( \frac{\eta_{\Delta}}{ \sigma_{0}}
    \right)^2 \right\} \\ \nonumber &+& A_\text{BE} \exp\left\{-
    \left[ \left( \frac{\phi_{\Delta}}{w_{\phi_{\Delta}}} \right)^2
      \hspace{-.08in} +\hspace{-.03in} \left(\frac{\eta_{\Delta}}{
          w_{\eta_{\Delta}}} \right)^2 \right]^{1/2} \right\}.~~
\end{eqnarray}
The definitions of two parameters in that expression ($A_D$ and $A_Q$)
are modified from those in Ref.~\cite{anomalous}.

Figure~\ref{fig1} (right panels) shows typical 2D model fits with
Eq.~\ref{2dmodel} compared to corresponding data histograms in the
left panels. The fit residuals are consistent with bin-wise
statistical errors. The general evolution with centrality is monotonic
increase of the SS 2D peak and AS dipole amplitudes (dijet structure),
substantial increase of the SS peak $\eta_\Delta$ width, rapid
decrease to zero of the 1D Gaussian on $\eta_\Delta$ (soft
component)~\cite{axialci,ptscale,anomalous} and non-monotonic
variation of the quadrupole amplitude~\cite{davehq}.

For the present 1D study we develop simplified versions of the
11-parameter model. In more-central \auau\ collisions the soft
component ($A_\text{soft}$) falls to zero amplitude, and the
BE/electron component ($A_\text{BE}$) becomes very
narrow~\cite{anomalous}. A 2D model applicable to more-central \auau\
collisions then has 6 parameters
\begin{eqnarray}
  A(\eta_\Delta,\phi_\Delta)   & = & 
  A_{2D}  \exp\left\{- \frac{1}{2} \left[ \left( \frac{\phi_{\Delta}}{ \sigma_{\phi_{\Delta}}} \right)^2 \hspace{-.05in} + \left( \frac{\eta_{\Delta}}{ \sigma_{\eta_{\Delta}}} \right)^2 \right] \right\} 
  \nonumber \\
  &+& A_{\rm D} [\cos(\phi_\Delta - \pi)+1]/2 
  \nonumber \\
  &+& A_0 
  + A_{\rm Q} 2\cos(2\, \phi_\Delta).
  \hspace{-.03in} 
\end{eqnarray}
The BE/electron component remains significant in a few bins near the
origin that can be removed from the fits.

Projection onto 1D azimuth represents large information reduction. The
full 2D histogram with $25\times 25$ bins includes 169 independent
bins (one independent quadrant due to symmetrization), whereas 1D
projections include at most 13 independent bins.
A simplified model derived from the 2D data model but applicable to
projected 1D azimuth correlations in more-central \aa\ collisions
includes 5 parameters defined to be consistent with the PS introduced
in Sec.~\ref{fourier}
\begin{eqnarray} \label{simpmod0} A(\phi_\Delta) & = & A_{1D}
  \exp\left[- \frac{1}{2} \left( \frac{\phi_{\Delta}}{
        \sigma_{\phi_{\Delta}}} \right)^2 \hspace{-.05in} \,\, \right]
  \nonumber \\
  &-& A'_{\rm D} 2\cos(\phi_\Delta)
  \nonumber \\
  &+& A_0 + A_{\rm Q} 2\cos(2\, \phi_\Delta).  \hspace{-.03in}
\end{eqnarray}

A further simplification is possible for the most-central (0-5\%)
bin. The quadrupole amplitude $A_Q$ for that centrality is observed to
be consistent with zero~\cite{davehq,davehq2}. The 1D model then
includes only 4 parameters
\begin{eqnarray} \label{simpmod} A(\phi_\Delta) & = & A_{1D}
  \exp\left[- \frac{1}{2} \left( \frac{\phi_{\Delta}}{
        \sigma_{\phi_{\Delta}}} \right)^2 \hspace{-.05in}\,\, \right]
  \nonumber \\
  &+& A_0 - A_{\rm D}' 2\cos(\phi_\Delta).
  % + A_0.
  \hspace{-.03in} 
\end{eqnarray}
Integrating Eqs.~(\ref{simpmod}) and (\ref{wiener1}) with differential
factor $d\phi_\Delta = 2\pi/N$ gives
\begin{eqnarray}
  2\pi P_0 &=& A_{1D} \sqrt{2\pi} \sigma_{\phi_\Delta} + 2\pi A_0~~~ \text{or}
  \\ \nonumber
  A_0 &=& - A_{1D} \sigma_{\phi_\Delta} / \sqrt{2\pi} + P_0.
\end{eqnarray}
The 1D data histograms have been adjusted to insure $P_0 \equiv 0$.  A
fit to bin-10 data with Eq.~\ref{simpmod} determines an offset value
$A_0 = -0.14$. With other fitted parameter values we obtain
\begin{eqnarray}
  A_{1D} \sigma_{\phi_\Delta} / \sqrt{2\pi} &=& 0.57 \times 0.635 / \sqrt{2\pi} = 0.144.
\end{eqnarray}
The four-parameter 1D model can then be further reduced to a
three-parameter model defined by
\begin{eqnarray} \label{simpmod2}
A(\phi_\Delta)   & = & 
A_{1D}\left\{  \exp\left[- \frac{1}{2} \left( \frac{\phi_{\Delta}}{ \sigma_{\phi_{\Delta}}} \right)^2 \hspace{-.05in}\,\,   \right] - \sigma_{\phi_{\Delta}} / \sqrt{2\pi} \right\}
 \nonumber \\
&-& A_{\rm D}' 2\cos(\phi_\Delta),
% + A_0.
\hspace{-.03in} 
\end{eqnarray}
where each of two model components integrates to zero over $2\pi$. We
therefore replace Eq.~(\ref{simpmod}) with Eq.~(\ref{simpmod2})
referred to below as the ``basic Model.''

Since all data histograms are corrected to $P_0 \equiv 0$ to remove
the offset DoF the adjusted 13-bin 1D data histograms have 12
independent DoF. But the bin at $\phi_\Delta = 0$ is removed from all
model fits to exclude the BE/electron component, reducing the
effective data DoF to 11.

The AS dipole component is the limiting case of an AS Gaussian peak
array (see App.~\ref{periodpeak} for details). The r.m.s.\ peak width
($\sigma \approx \pi/2$) is large enough that only the $m = 1$ AS
dipole term of the PS representation survives.

We define alternative data models by adding to the basic Model of
Eq.~(\ref{simpmod2}) successive cosine terms of the form $A_X 2 \cos(m
\phi_\Delta)$, where $X = Q,~S,~O$ for quadrupole, sextupole and
octupole ($m = 2,~3,~4$). We also define independent ``FS-only''
models as truncated Fourier series with $K$ cosine terms and no other
components.

%%%%%%%%%%%%%%%%%%%%
\section{Bin-10 0-5\% Azimuth correlations} \label{bin10a}

We first apply BI methods to the 1D azimuth projection from 0-5\%
central 200 GeV \auau\ collisions.
We fit the data with the basic Model and obtain the data PS. We
determine $\chi^2$ and information $I$ for FS-only models vs parameter
number $K$. We then evaluate evidence $E$ for several competing models
and determine the posterior model probabilities.

\subsection{1D azimuth projection} \label{1dazfit}

\begin{figure}[h]
\includegraphics[width=.48\textwidth]{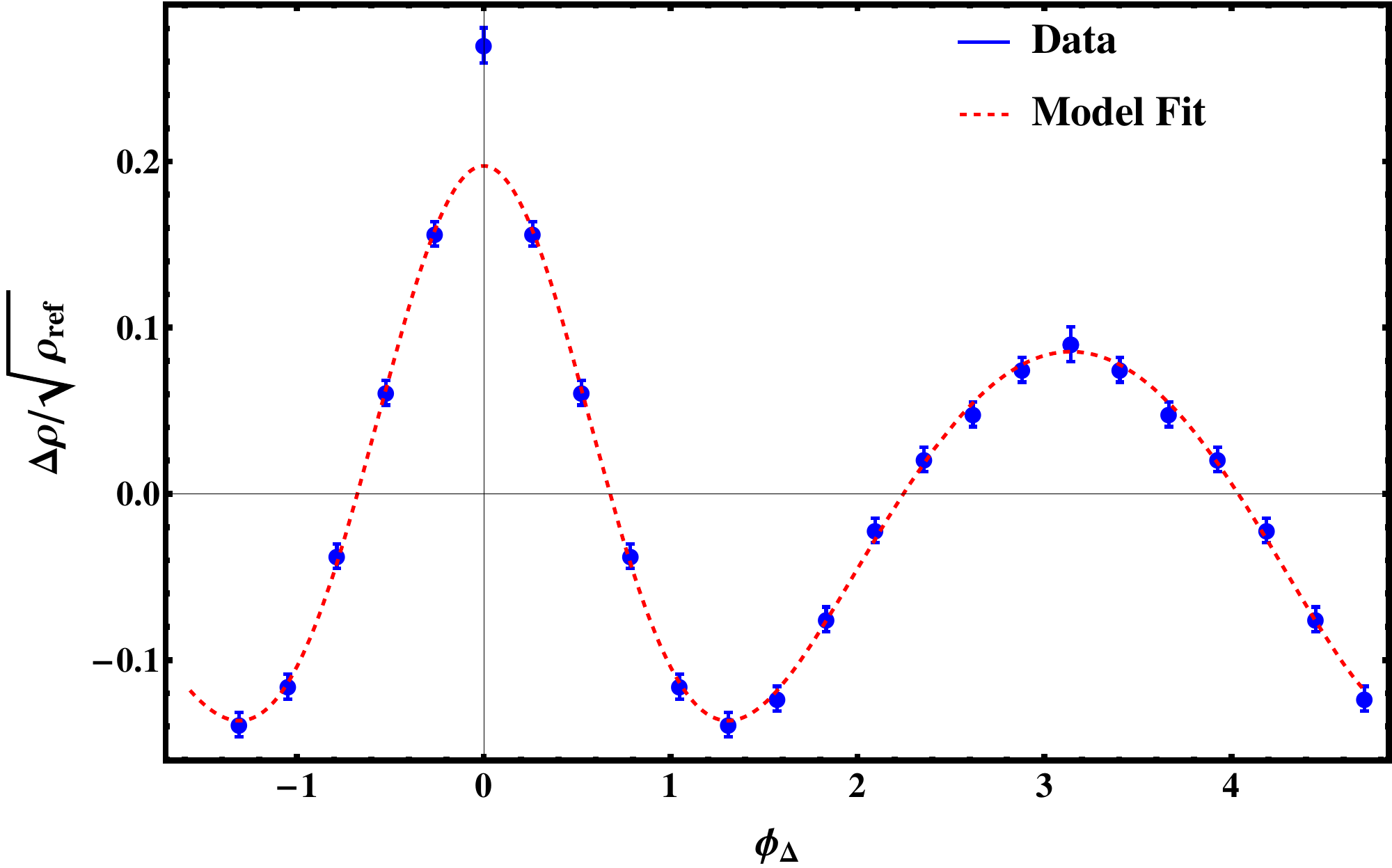}   
% \put(-28,90) {\bf (a)} \put(-158,120) {\bf bin 10}
\caption{\label{1ddata} (Color online) 1D projection onto azimuth
  (points) from the 2D data histogram for 0-5\% central 200 GeV \auau\
  collisions in Fig.~\ref{fig1} (c). Statistical errors at 0 and $\pi$
  are $\sqrt{2}$ larger than the others due to symmetrization of data
  on the periodic variable. The bin-wise statistical errors 0.0037
  have been multiplied by 2 to make them visible outside the data
  points. The (red) dashed curve is obtained from a fit to the data
  with the basic Model of Eq.~(\ref{simpmod2}). A fit with an FS-only
  model including four or more terms would appear identical on the
  scale of this plot. A similar remark applies to corresponding data
  plots for two other centrality bins.
} %AuAu2002GeV-bin10-Projection
\end{figure}

Figure~\ref{1ddata} shows a projection of the 2D data histogram from
0-5\% central 200 GeV \auau\ collisions onto 1D $\phi_\Delta$
(points). 24 bins are shown but only 13 are unique due to
symmetrization about zero and $\pi$. Estimated statistical errors have
been multiplied by factor 2 to make them visible (extend outside the
points). Errors are a factor $\sqrt{2}$ larger for the bins at 0 and
$\pi$ because of symmetrization of the data about those bins. The bin
at zero also includes a significant contribution from BE/electrons not
included in the models used for this exercise and is therefore
excluded from all fits. The bin at $\pi$ includes a small excess due
to a tracking-geometry distortion accommodated in some model fits by
addition of a ``delta function.''

%%%%%%%%%%%%%%%%%%%%%%%
%%%%%%%%%%%%%%%%%%%%%%%

A fit of the basic Model to data is shown by the dashed (red)
curve. The fitted model parameters are $A_{1D} =0.57\pm 0.007$,
$\sigma_{\phi_\Delta} =0.635\pm 0.007$ and $A'_D = 0.115\pm 0.002$
with $\chi^2 = 12.5$ for $11 - 3 = 8$ fit DoF.

\subsection{Data power spectrum} \label{powerspec}

\begin{figure}[h]
%\put(-28,90) {\bf (a)}
\includegraphics[width=.48\textwidth]{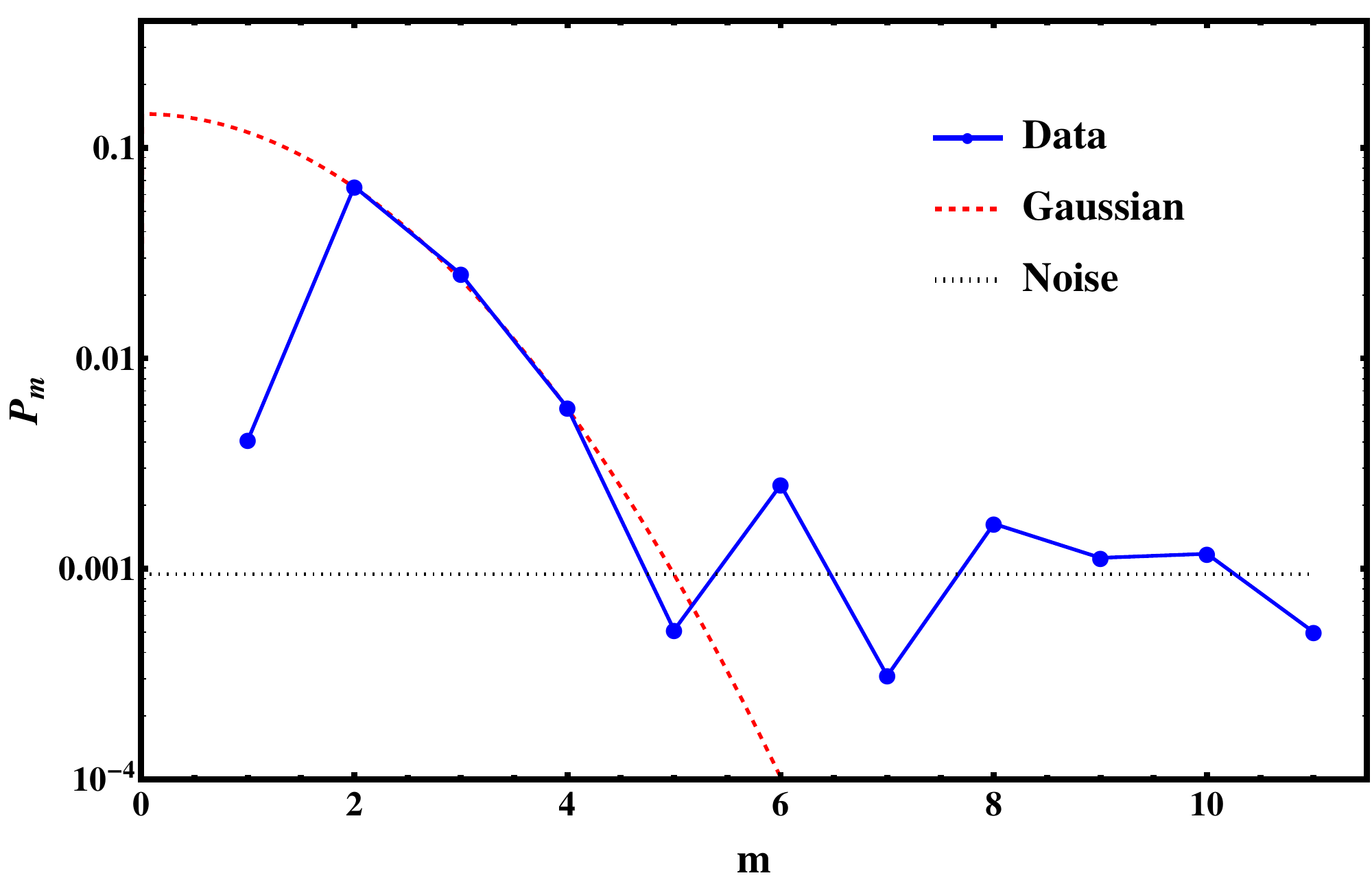}   
% \put(-158,120) {\bf bin 10}
\caption{\label{power1} (Color online) Power spectrum values $P_m$
  (points) derived from the data in Fig.~\ref{1ddata} via
  Eq.~(\ref{coef}). The (red) dashed curve is the Gaussian PS
  described by Eq.~(\ref{fmm}) with width and amplitude corresponding
  to the fitted Gaussian in Fig.~\ref{1ddata}. Interval $m \geq 5$ is
  consistent with a ``white-noise'' power spectrum (dotted line)
  representing the statistical noise in Fig.~\ref{1ddata}.
} %AuAu200GeV-bin10-PowerSeries
\end{figure}

Figure~\ref{power1} shows the PS (points and blue solid curve) as a
Fourier transform of the data autocorrelation in Fig.~\ref{1ddata}
using Eq.~(\ref{coef}). The general structure includes a signal
component at smaller wave number $m \leq 4$ and a flat (on average)
white-noise spectrum at larger wave number corresponding to the
r.m.s.\ statistical error in the data histogram. The noise-spectrum
mean is about 0.001 (dotted line).

%%%%%%%%%%%%%%%%%%%%%%%
%%%%%%%%%%%%%%%%%%%%%%%

To aid interpretation of the data PS we include the predicted PS for a 1D Gaussian (red dashed curve) with amplitude and width derived from the basic-Model fit  in Fig.~\ref{1ddata}. The PS amplitudes for a {\em unit-amplitude} periodic Gaussian peak array on $\phi_\Delta$ are given by (App.~\ref{periodpeak})
\begin{eqnarray} \label{fmm} 2P_m(\sigma_{\phi_\Delta}) &=&
  \sqrt{2/\pi}\, \sigma_{\phi_\Delta} \exp\left( - m^2
    \sigma_{\phi_\Delta}^2 / 2 \right).
\end{eqnarray}
As the Gaussian peak width $\sigma_{\phi_\Delta}$ increases the number
of significant signal terms in the PS decreases.
The Gaussian PS coincides with the data PS for $m \in [2,5]$, and the
data PS for $m \geq 5$ is consistent with statistical noise. The data
PS element for $m = 1$ includes a negative contribution $-A'_D =
-0.115$ from the AS peak (dipole).

We can assess the quality of the basic-Model data description by
determining the PS of the residuals, not of (data $-$ Model) but of
(data $-$ Gaussian) only. The PS of the residuals should be equal to
the PS difference in Fig.~\ref{power1} according to the linearity of
Eq.~(\ref{coef}).

%%%%%%%%%%%%%%%%%%%%%%%
\begin{figure}[h]
\includegraphics[width=.48\textwidth]{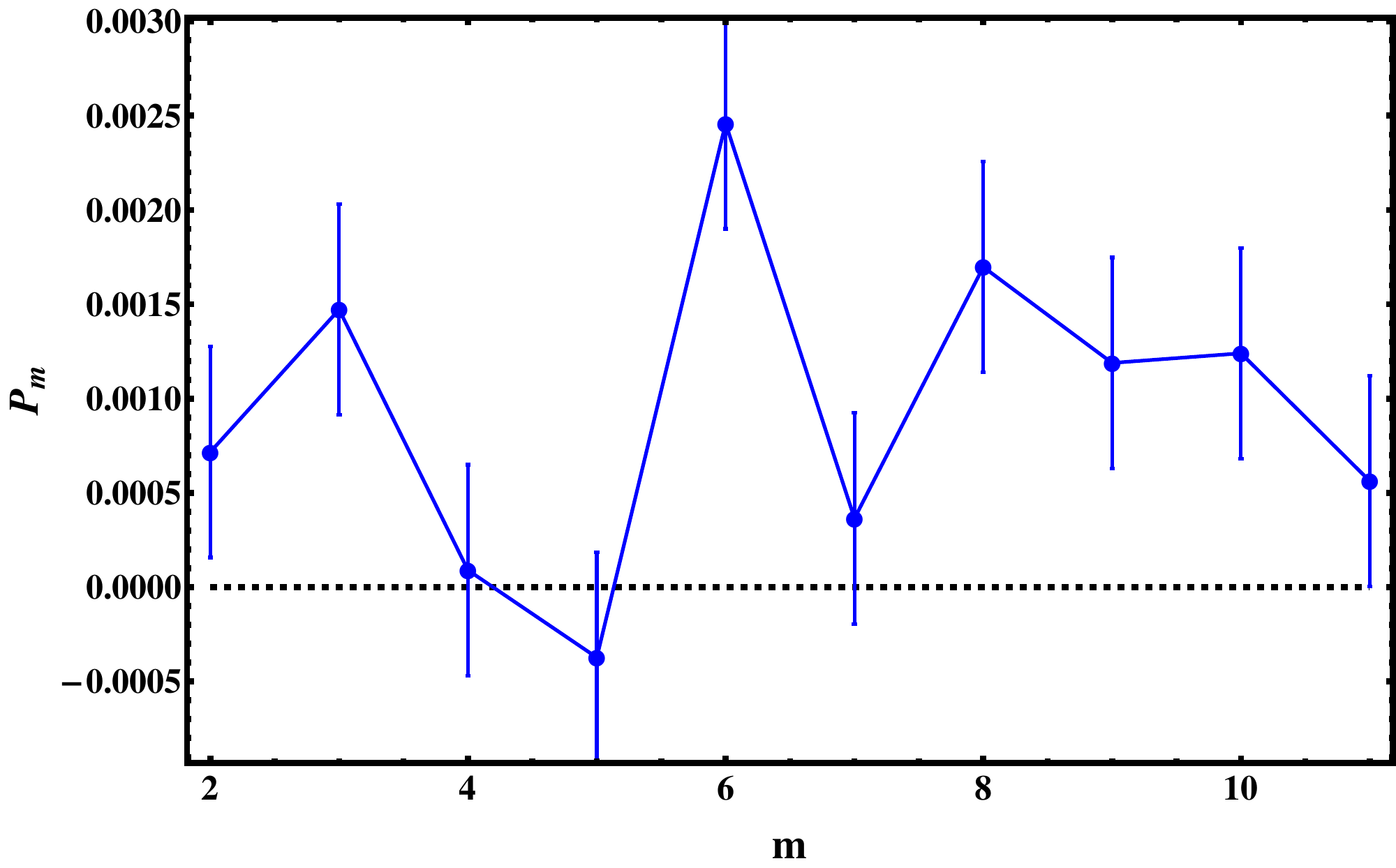}   
%\put(-158,120) {\bf bin 10}
\caption{\label{noise} (Color online) The PS for residuals in the form
  (data $-$ Gaussian) from Fig.~\ref{1ddata} consistent with the
  white-noise part of the PS in Fig.~\ref{power1}, with mean
  approximately 0.001. The negative PS value for $m = 1$ (not shown)
  corresponds to the AS dipole amplitude $-A'_{D}$ from the
  basic-Model fit in Fig.~\ref{1ddata}.
} %AuAu200GeV-bin10-PowerResidue
\end{figure}
%%%%%%%%%%%%%%%%%%%%%%%

Figure~\ref{noise} shows the PS for (data $-$ Gaussian) referring to
the fitted Gaussian in Fig.~\ref{1ddata}. The PS values for $m > 1$
are consistent with the white-noise spectrum. The value for $m = 1$
(not shown) is consistent with the fitted dipole amplitude. From this
PS study we have a first indication that the $K = 3$ basic Model is
{\em sufficient} to describe the bin-10 1D azimuth projection.

\subsection{Bayesian model fits with Fourier series}

We next apply BI methods to FS-only models of the data histogram in
Fig.~\ref{1ddata} to establish a BI reference. In this application the
number of parameters $K$ represents the largest value of FS index $k$
for a given FS-only model. Varying $K$ represents different FS data
models.  We obtain the $\chi^2$ and information $I$ for each FS-only
model.

%%%%%%%%%%%%%%%%%%%%%%%
\begin{figure}[h]  
\includegraphics[width=.48\textwidth]{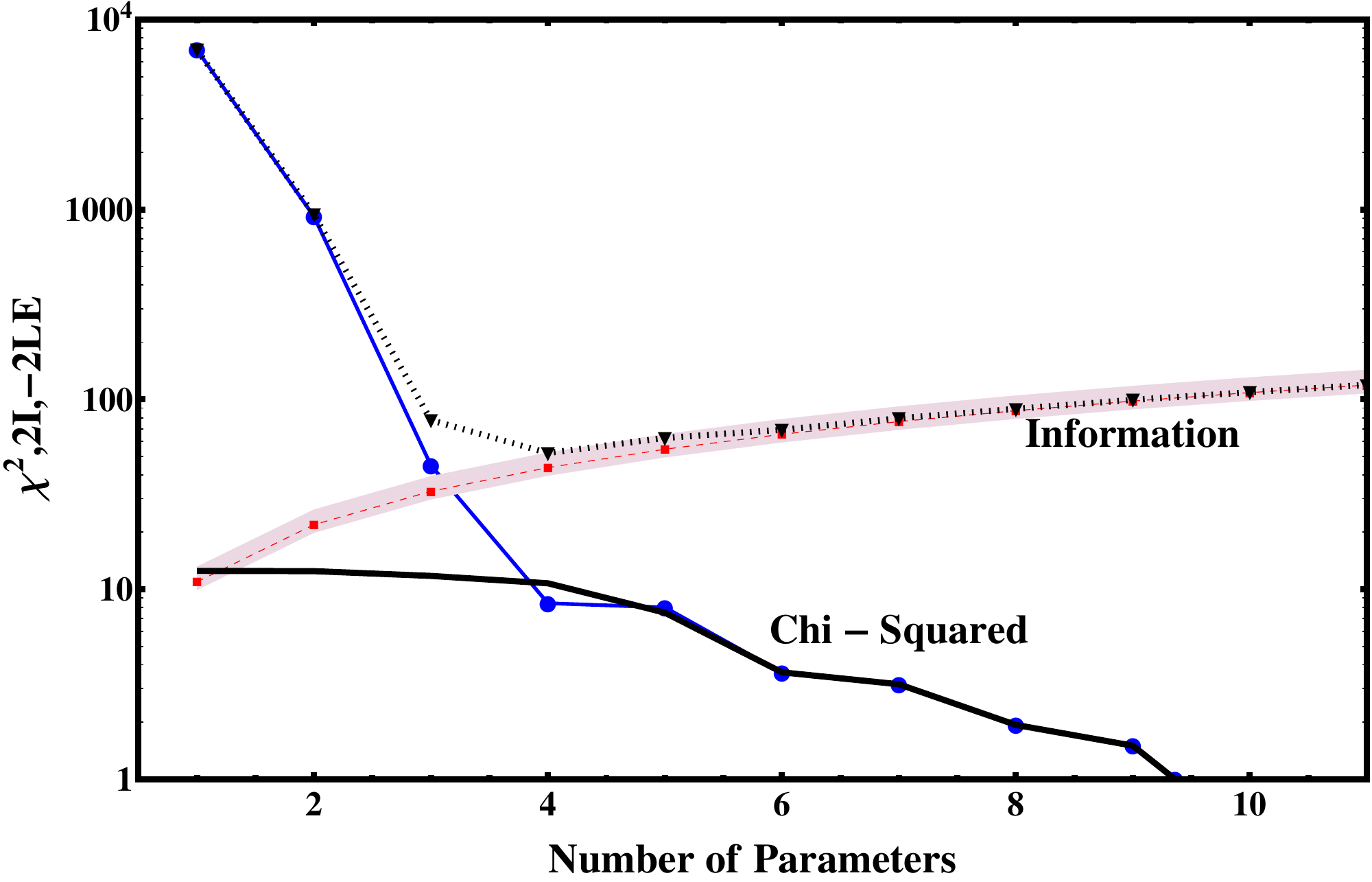}   
%\put(-28,90) {\bf (a)}
%\put(-158,120) {\bf bin 10}
\caption{\label{basicfit} (Color online) $\chi^2$ (upper solid curve
  and points) and information $2I$ (dashed curve and points with
  uncertainty band) vs number of parameters $K$ for Fourier-series
  (FS-only) models. The sum (log Evidence, $-2LE$, dotted curve) is
  also included. The lower solid curve is $\chi^2$ values for fits to
  residuals (data $-$ basic Model) from Fig.~\ref{1ddata} consistent
  with the trend $11 - K$ expected for no signal (noise only) in the
  data.  } %AuAu200GeV-bin10-ChiSquared
\end{figure}
%%%%%%%%%%%%%%%%%%%%%%%

Figure~\ref{basicfit} shows the basic elements of BI model fits. The
upper (blue) solid curve and points represent the log likelihood (LL)
in the form $-2LL$ or $\chi^2$. The (red) dashed curve shows
information $2I$ representing the parameter cost (Occam penalty,
Sec.~\ref{occam}). The (black) dotted curve represents the sum $-2LE$
(negative log evidence). The minimum for $-2LE$ (and maximum for
evidence $E$) occurs at $K = 4$ indicating the FS-only model preferred
by the data. That result is consistent with Fig.~\ref{power1}
indicating a $K = 4$ FS-only model should exhaust the PS signal.

The $\chi^2$ trend indicates that the FS model components are ideally
ordered on index $k \in [1,K]$ for the signal in these specific data,
and is similar to the idealized trend suggested in Fig.~5.1 of
Ref.~\cite{bayes2}. The largest decreases occur for the smallest index
values. The interval with larger (negative) slope at smaller $K$
corresponds to accommodation of the data signal with increasing
$K$. The interval with smaller slope at larger $K$ indicates that
additional Fourier terms only accommodate statistical noise. The
overall $\chi^2$ trend then matches the power-spectrum trend in
Fig.~\ref{power1}. $\chi^2$ must go to zero when $K =$ the number of
data DoF (11 in this case). The lower solid curve represents the
$\chi^2$ for fits to the residuals (data $-$ basic Model) from
Fig.~\ref{1ddata} (no signal present). The $\chi^2$ values are then
consistent with the fit DoF $\approx 11 - K$.

\subsection{Bayesian model comparisons} \label{modcomp}

We next extend BI methods to several data models with different
combinations of elements and parameters compared to the previous
FS-only exercise. We first compare $\chi^2$ alone, simulating a
conventional model-fit exercise, then extend to comparisons of
evidence $E(D^*|H)$.

%%%%%%%%%%%%%%%%%%%%%%%
\begin{figure}[h]
\includegraphics[width=.48\textwidth]{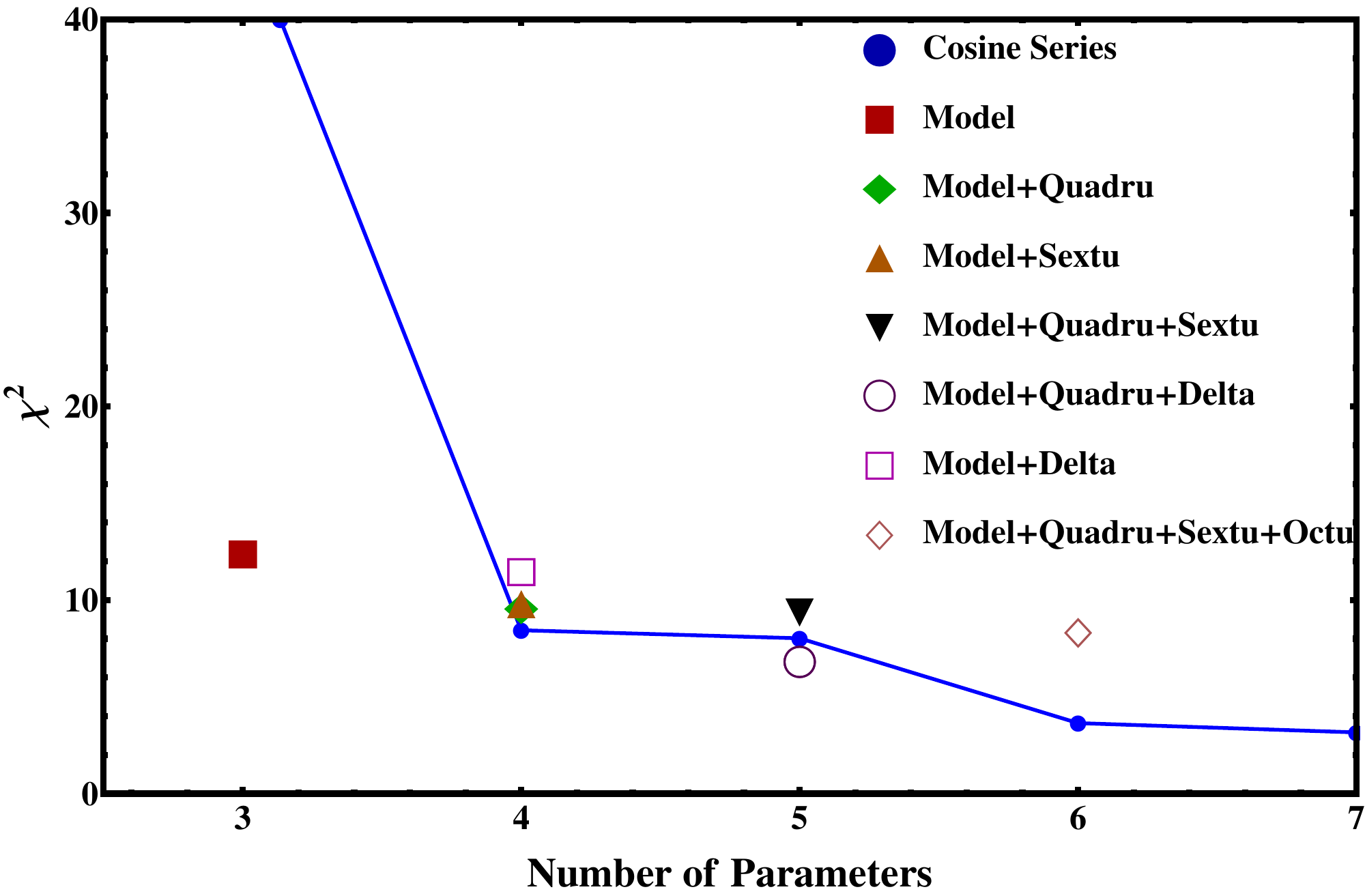}   
%\put(-28,90) {\bf (a)}
%\put(-158,120) {\bf bin 10}
\caption{\label{chi2} (Color online) $\chi^2$ values vs number of
  parameters $K$ for several data models. The general trend is
  monotonic decrease with increasing number of model parameters,
  responding only to statistical noise with $\chi^2 \approx 11 - K$
  for $K > 4$.  } %AuAu200GeV-bin10-Chi2Modelsy
\end{figure}
%%%%%%%%%%%%%%%%%%%%%%%

Figure~\ref{chi2} shows $\chi^2$ values for various model fits to
data. The FS-only description (blue points and line) achieves a
substantial decrease for $K = 4$ but no significant improvement with
additional terms. The basic Model with three parameters (red solid
square) has $\chi^2 \approx 12.5$, somewhat in excess of the number of
fit DoF $= 8$. Addition of more cosines (quadrupole, sextupole,
octupole) to the basic Model keeps pace with the FS noise trend with
its reduced slope. In this conventional context the extra cosines {\em
  seem} to be required for competitive data description because they
reduce the fitted $\chi^2$, but at what cost?

The basic Model + quadrupole + sextupole + octupole with $K = 6$ (open
diamond) has the same $\chi^2$ as the $K = 4$ FS-only model. As
explained below in connection with Table~\ref{paramx} the additional
cosine terms effectively displace the Gaussian part of the basic
Model. The composite model then functions as a FS-only model with $K =
4$, but with increased cost in the Occam penalty.

%%%%%%%%%%%%%%%%%%%%%%%
\begin{figure}[h]
\includegraphics[width=.48\textwidth,height=.3\textwidth]{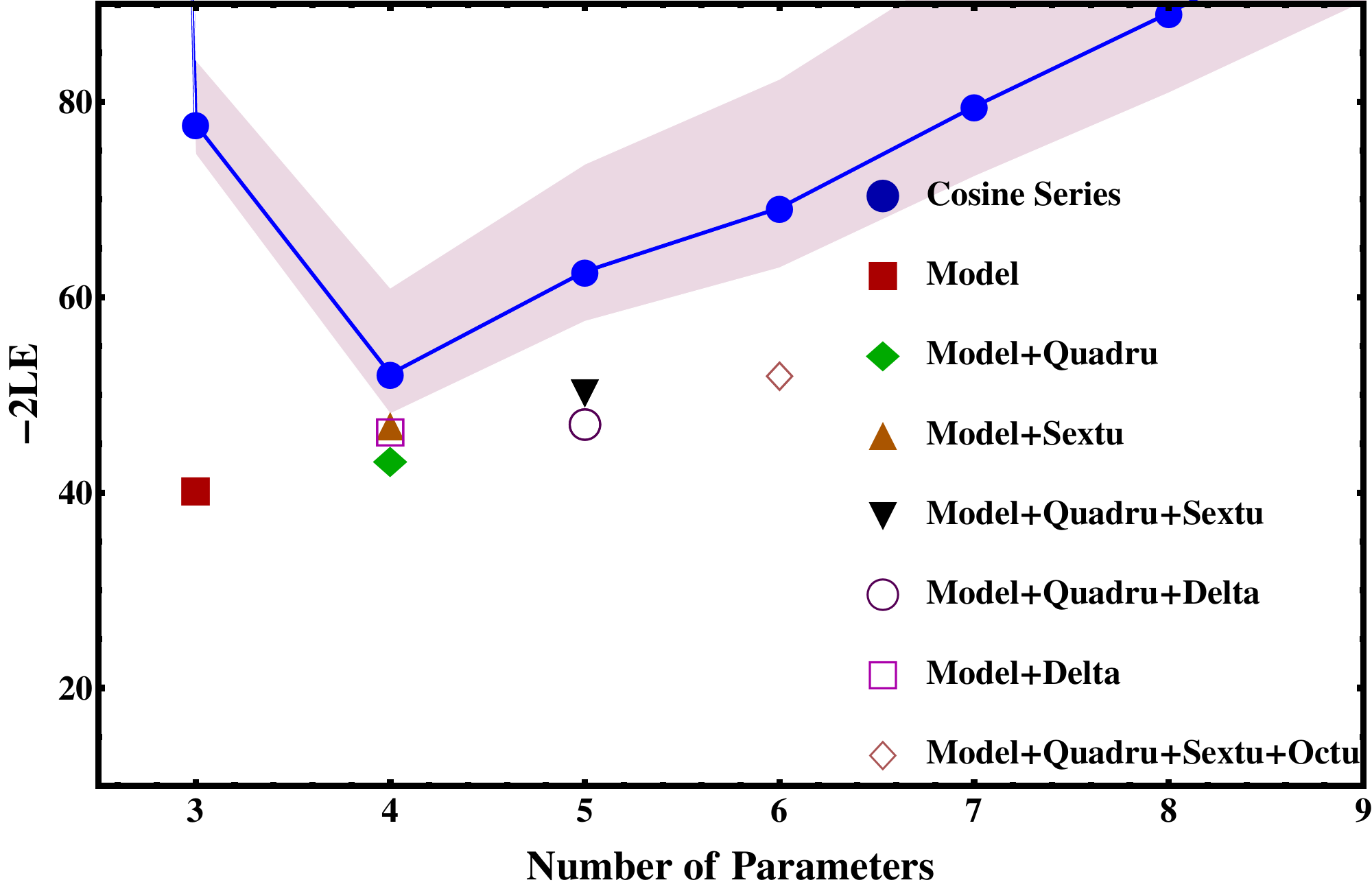}   
% \put(-28,90) {\bf (a)} \put(-158,120) {\bf bin 10}
\caption{\label{modelcomp} (Color online) Negative log evidence $-2LE$
  vs number of parameters $K$ for several models. The basic Model
  (solid square) is strongly favored over all others (lowest
  $-LE$). The hatched band indicates the common uncertainty of priors
  assigned to cosine terms in all models. FS-only models for all $K$
  (solid dots and line) are strongly rejected by the evidence.
} %AuAu200GeV-bin10-EvidenceModels
\end{figure}
%%%%%%%%%%%%%%%%%%%%%%%

Figure~\ref{modelcomp} shows negative log evidence $-2LE = \chi^2 +
2I$ for several models. Adding an Occam penalty in the form of
information $I$ gained by each model reveals a different picture. The
basic Model with $K = 3$ has substantially smaller -2LE (larger
evidence $E$) than other models where the cost of extra parameters is
not justified by a compensating reduction in $\chi^2$. The hatched
band reflects the estimated uncertainty in $I$ (for the FS-only model)
arising from the estimated priors.

Given that $\chi^2$ values for various models are similar ($\approx 11
- K$) the large differences in $-LE$ among models must be dominated by
information $I$ which depends on the covariance matrix and prior PDFs.
It might be suggested that such differences arise mainly from the
assignment of prior probabilities, but that is not the case.  We apply
the same prior to a given parameter or parameter class consistently
across all models, so that uncertainties in I are strongly correlated
across competing models and largely cancel when odds ratios are taken
(see Sec.~\ref{oddsys}).

The $LE$ trend vs $K$ for FS-only models arises from $2I \approx 10K$,
whereas the $LE$ trend for the basic Model plus additional cosine
terms corresponds to $2I \approx 5K$. The difference in $I/K$ of 2.5
corresponds to a factor $\exp(2.5) \approx 12$ difference in parameter
errors for the two models. Parameter errors for FS-only models are
$O(0.001)$ whereas errors for the basic Model plus cosine terms are
$O(0.01)$, accounting for the factor 10-15 difference. As discussed in
Sec.~\ref{why} the large Occam penalty for FS-only models is mainly
owing to smaller covariance-matrix elements (parameter errors).

%%%%%%%%%%%%%%%%%%%%%%%
\begin{figure}[h]  
\includegraphics[width=.48\textwidth,height=.3\textwidth]{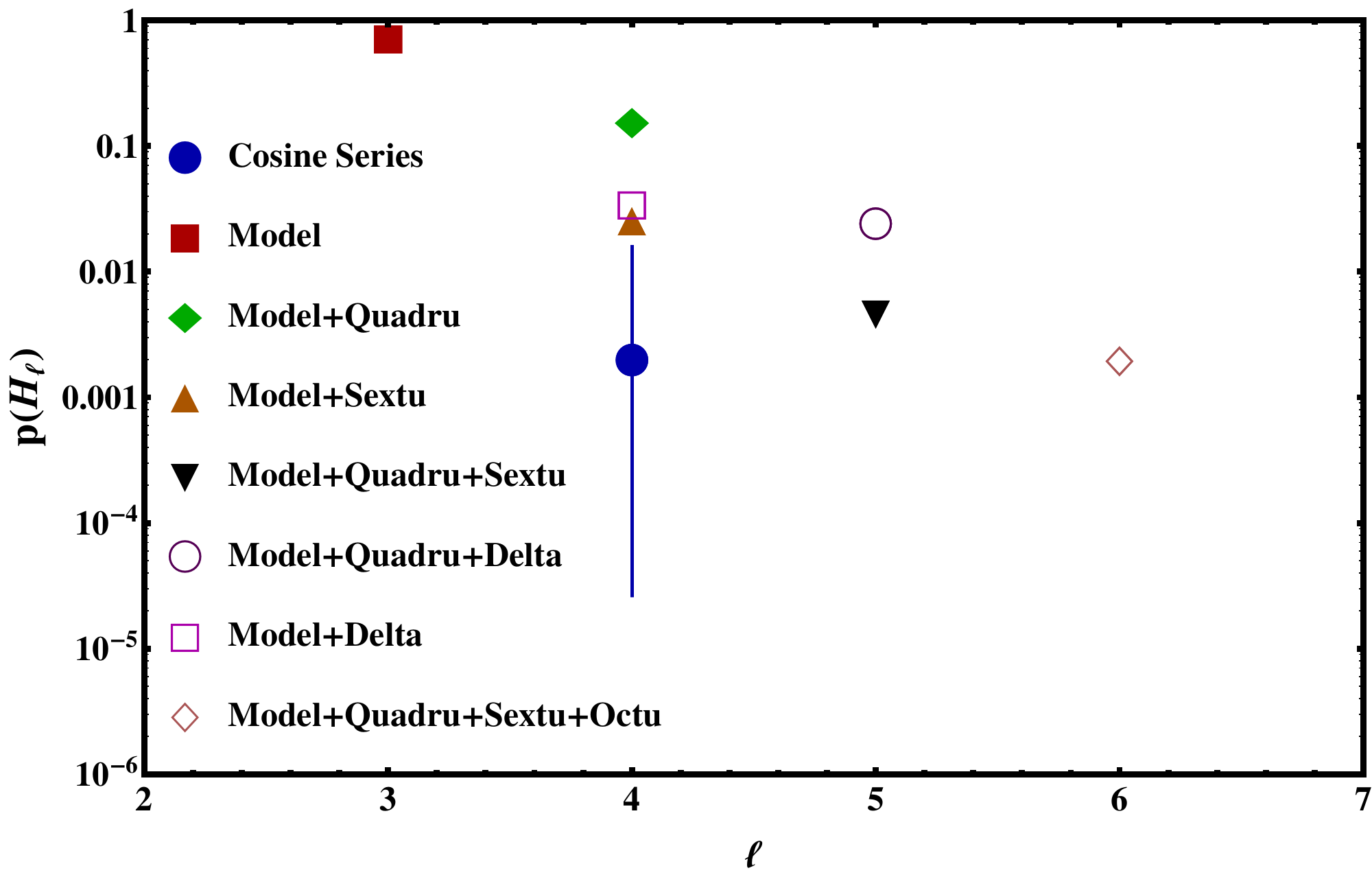}   
%\put(-28,90) {\bf (a)}
%\put(-158,120) {\bf bin 10}
\caption{\label{xxxx} (Color online) Normalized $p(H_l|D^*)$ from
  Eq.~(\ref{plaus}) for several models indexed by $l$. As in
  Fig.~\ref{modelcomp} the basic Model (solid square) is strongly
  favored over all other models while FS-only models for all $K$ are
  strongly rejected.  } %AuAu200GeV-bin10-ModelPosterior
\end{figure}
%%%%%%%%%%%%%%%%%%%%%%%

Figure~\ref{xxxx} shows the {\em plausibility} (relative evidence) for
each competing model in the form
\begin{eqnarray} \label{plaus}
p(H_l|D^*) &=& \frac{E(D^*|H_l) p(H_l)}{\sum_l E(D^*|H_l) p(H_l)}
\end{eqnarray}
that reveals the full selectivity of the BI method. For this exercise
we assume that model prior probabilities $p(H_l)$ are all equal (and
therefore irrelevant). However, implicit $p(H_l)$ assumptions do play
a role in RHIC/LHC data modeling and physics interpretations.

The most plausible models are the basic Model (80\%) and basic Model +
quadrupole (15\%). Large Occam penalties reduce competing additional
multipole elements to a few percent or less. The model including an
octupole (open diamond) leads to major fit instabilities and is
rejected. With plausibilities of less than 1\% FS-only models are also
rejected. In terms of odds the basic Model is preferred over Model +
quadrupole by $4.6 \pm 0.7$:1, over Model + sextupole by $28.0 \pm
4.8$:1 and over all FS-only models by $360 \pm 42$:1.

As noted in Section II-E Bayesian comparisons among models are
effected by taking ratios of evidences (odds ratios). Comparisons are
visualized efficiently by corresponding differences on a log-evidence
scale (Bayes factors) as in Fig.~\ref{modelcomp} and subsequent
equivalent figures. Isolated absolute numbers are not relevant to our
method.

\subsection{Model-fit results for bin 10}

Table~\ref{paramx} summarizes the best-fit model-parameter values
$\tilde w$ obtained from model fits (minimum $\chi^2$) emphasizing the
basic Model (column 2) and successive additions of quadrupole,
sextupole and octupole components, as well as a delta function at
$\pi$ to accommodate a data artifact. The parameters are as defined in
Sec.~\ref{models}. Also shown are $\chi^2$ and the BI parameters $2I$
and $-2LE$.

%%%%%%%%%%%%%%%%%%%%%%%%%%%%%%%%%%
\begin{table}[h]
  \caption{  \label{paramx}
    Bin-10 model parameters (minimum $\chi^2$) for several fit models: 
    (a)  basic Model,
    (b) basic Model plus quadrupole term $A_Q$, 
    (c)  previous  plus sextupole term $A_S$,
    (d)  previous  plus octupole term $A_O$,
    (e) basic Model plus delta function at $\pi$.
    The fit parameters are as defined in Sec.~\ref{models}. 
  }
%%%%%%%%%%%%%%%%%%%%%%%%%%%%%%%
\begin{center}
\begin{tabular}{|c|c|c|c|c|c|} \hline
%\multicolumn{8}{|c|c|}{ multiplicity classes} \\ \hline
 parameter & basic Model &  + $A_\text{Q}$ &  + $A_\text{S}$ &  + $A_\text{O}$ & $+ \delta$ \\ \hline \hline
$A_{1D}$                    & 0.57$\pm0.007$    & 0.73$\pm0.09$ & 0.84& 0.34 & 0.57  \\ \hline
 $\sigma_{\phi_\Delta}$ & 0.64$\pm 0.007$      & 0.69$\pm0.02$ & 0.71 & 0.09 & 0.63  \\ \hline
%$A_0$                         & -0.14$\pm 0.0$   & -0.20 & -0.24  & 0.002 &  -0.14 \\ \hline
$A'_D$                       & 0.12$\pm 0.002$    & 0.15$\pm0.02$ & 0.18  & -0.003 & 0.115  \\ \hline
$A_Q$                       & --                &-0.014$\pm 0.007$   & -0.025 & 0.064 &  -- \\ \hline
$A_S$                       & --                & --               & 0.005  & 0.024 & --  \\ \hline
$A_O$                      & --                 & --               & --    & 0.005 &  --  \\ \hline 
$A_\delta$                      & --                 & --               & --    & -- &  0.005  \\ \hline   \hline
$\chi^2$                    & 12.5            & 9.7             & 10 & 9 &  11  \\ \hline
$2I$                          & 28              & 34              & 38 & 44 &  36  \\ \hline
$-2LE$                     & 40.5            & 42.5              & 48 & 53 &  47  \\ \hline
%$-2LE$                     & 52.5            & 56              & 61 & 57 &  61  \\ \hline
\end{tabular}
\end{center}
%%%%%%%%%%%%%%%%%%%%%%%%%%%%%%%
\end{table}
%%%%%%%%%%%%%%%%%%%%%%%%%%%%%%%%%%

Results for the basic Model are in good agreement with the published
values from 2D model fits~\cite{anomalous}. For this centrality the
best-fit 2D parameters from the model of Eq.~(\ref{2dmodel}) are
$A_{2D} = 0.65\pm0.04$, $\sigma_{\phi_\Delta} = 0.63\pm0.015$, $A_0 =
-0.14\pm0.014$ (consistent with the Gaussian integral), $A_D =
0.224\pm0.002$ ($\approx 2A'_D$), $A_Q = 0.001\pm0.008$ with $\chi^2$
/ DoF = 2.6. Note that $A_{1D}$ must be less than $A_{2D}$ because of
the curvature on $\eta_\Delta$ of the SS 2D peak. The $\chi^2$/DoF =
2.8 of the 2D model fit is substantially higher than that for the 1D
fit with the basic Model [12.5 / (11 - 3) = 1.6] because of
significant structure on $\eta_\Delta$ ($\eta$-modulated dipole) not
described by the standard 2D data model of Eq.~(\ref{2dmodel}).

As cosine terms are added to the basic Model a conflict develops
between the explicit Gaussian component and a sum of cosines
approximating a competing Gaussian.  The large parameter differences
for ``$+A_O$'' vs ``basic Model'' columns are discussed further in
Sec.~\ref{compete}.

The $+\delta$ column refers to the basic Model plus a free amplitude
in the bin at $\pi$ (``delta function''). Compared to the basic Model
alone there is reduction of $\chi^2$ by 1.5 but increase of
information $2I$ by 8 leading to overall increase of negative log
evidence $-2LE$ by 6.5. The additional model DoF is rejected by
$\exp(3.25) \rightarrow$ 25:1.

%%%%%%%%%%%%
\section{bin-8 9-18\% Azimuth correlations} \label{bin8a}

In this second of three examples the statistical errors of the wider
centrality bin are reduced by factor $\sqrt{2}$ compared to the 0-5\%
centrality bin. The BE/electron peak is still narrow enough to remain
within the single bin at zero. The quadrupole component is significant
and positive, shifting the plausibility order of competing models.

\subsection{1D azimuth projection}

\begin{figure}[h]  
  \includegraphics[width=.48\textwidth]{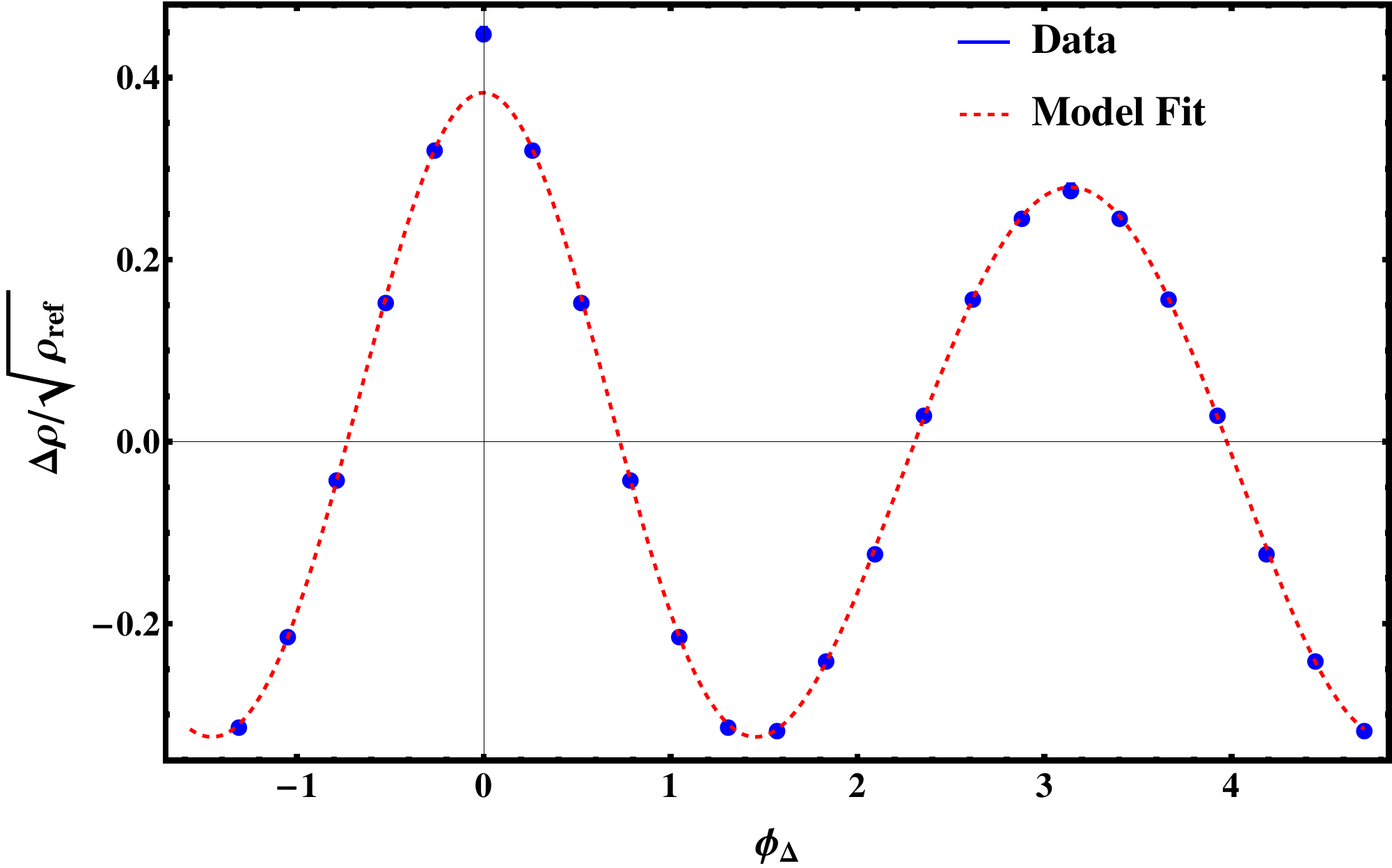}
  \caption{\label{1ddata2} (Color online) 1D projection onto azimuth
    (points) from the 2D data histogram for 9 - 18\% central 200 GeV
    \auau\ collisions. The (red) dashed curve is a fit to the data
    with the basic Model of Eq.~(\ref{simpmod2}) plus independent
    quadrupole component $A_Q 2 \cos(2 \phi_\Delta)$. The bin-wise
    statistical errors are 0.0026, not visible outside the points.
  } %AuAu2002GeV-bin8-Projection
\end{figure}

Figure~\ref{1ddata2} shows a projection of the 2D data histogram from
9-18\% central 200 GeV \auau\ collisions onto 1D $\phi_\Delta$
(points). As for the previous centrality the bin at zero also includes
a significant contribution from BE/electrons not included in the data
models and is therefore excluded from the fits.  The typical data
r.m.s.\ statistical error is 0.0026, not visible outside the points on
this scale.

%%%%%%%%%%%%%%%%%%%%%%%
%%%%%%%%%%%%%%%%%%%%%%%

A fit of the basic Model + quadrupole to data is shown by the dashed
(red) curve. The fitted model parameters are $A_{1D} =0.926\pm0.088$,
$\sigma_{\phi_\Delta} =0.727\pm0.018$, $A'_D = 0.206\pm0.022$ and $A_Q
= 0.068\pm0.006$ with $\chi^2 = 16.5$ for 11 - 4 = 7 fit DoF.

\subsection{Data power spectrum} \label{powerspec3}

\begin{figure}[h]  
%\put(-28,90) {\bf (a)}
\includegraphics[width=.48\textwidth]{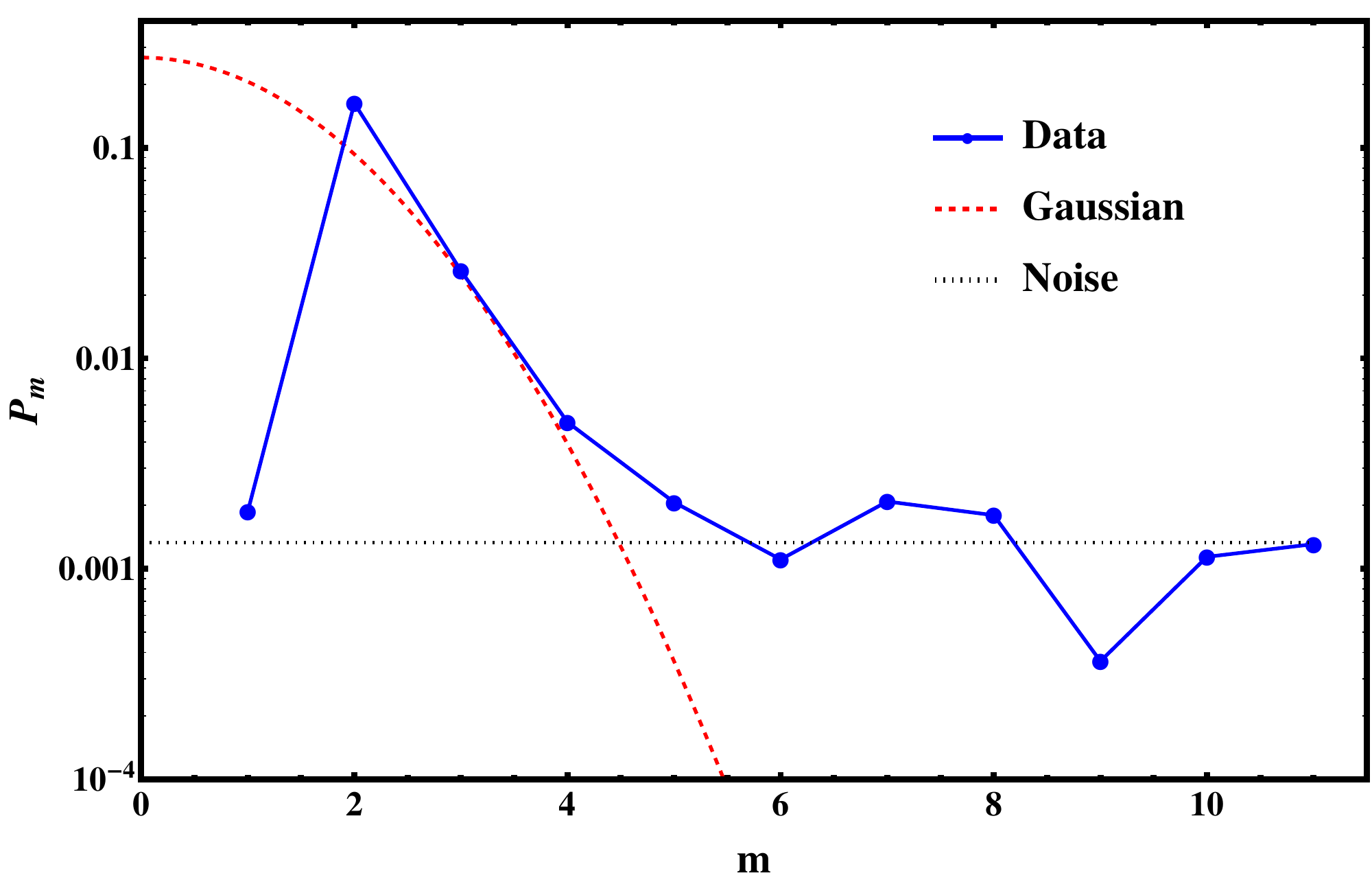}   
% \put(-158,120) {\bf bin 10}
\caption{\label{power2} (Color online) Power spectrum values $P_m$
  (points) derived from the data in Fig.~\ref{1ddata2} via
  Eq.~(\ref{coef}). The (red) dashed curve is the Gaussian PS
  described by Eq.~(\ref{fmm}) with amplitude and width from the basic
  Model + quadrupole fitted to data in Fig.~\ref{1ddata2}. The
  interval $m \geq 5$ is consistent with a ``white-noise'' power
  spectrum (dotted line) representing the statistical noise in
  Fig.~\ref{1ddata2}.  } %AuAu200GeV-bin8-PowerSeries
\end{figure}

Figure~\ref{power2} shows the power spectrum (points and blue solid
curve) derived from the data in Fig.~\ref{1ddata2}.
As for the 0-5\% centrality bin we include the predicted power
spectrum (red dashed curve) for a 1D Gaussian (SS peak) with amplitude
and width parameters derived from the fit to data in
Fig.~\ref{1ddata2}. The data PS is again consistent with statistical
noise for $m \geq 5$. The PS element for $m = 1$ includes a negative
contribution from the AS dipole.  The element for $m = 2$ includes a
significant positive contribution from a quadrupole component not
associated with the SS peak~\cite{davehq}.

%%%%%%%%%%%%%%%%%%%%%%%
%%%%%%%%%%%%%%%%%%%%%%%

Just as for bin 10 we assess the quality of the basic-Model data
description by determining the PS of the residuals of (data $-$
Gaussian) only, where Gaussian is the fitted Gaussian in
Fig.~\ref{1ddata2}. The PS for (data $-$ Gaussian) is consistent with
a white-noise spectrum with mean value $\approx 0.0013$ for $m > 2$.
The values for $m = 1,$ 2 are consistent with the fitted positive
quadrupole and negative dipole amplitudes.
The basic Model augmented by quadrupole component $\cos(2\phi_\Delta)$
fully exhausts the data signal and is therefore a sufficient model.

\subsection{Bayesian model fits with Fourier series}

The log-likelihood LL trend in the form $\chi^2$ for $K > 2$ for
FS-only model fits to data from bin 8 (not shown) is similar to that
for bin-10 data in Fig.~\ref{basicfit}. Information $2I$ representing
the parameter cost is also similar. The minimum of $-2LE$ occurs at $K
= 4$, consistent with Fig.~\ref{power2} where we again find that a $K
= 4$ FS-only model should completely describe the signal in the bin-8
data. The FS-only model should then be competitive with the $K = 4$
basic Model + quadrupole in terms of fit quality and parameter number,
two elements of BI evaluation.

\subsection{Bayesian model comparisons}

\begin{figure}[h]  
\includegraphics[width=.48\textwidth]{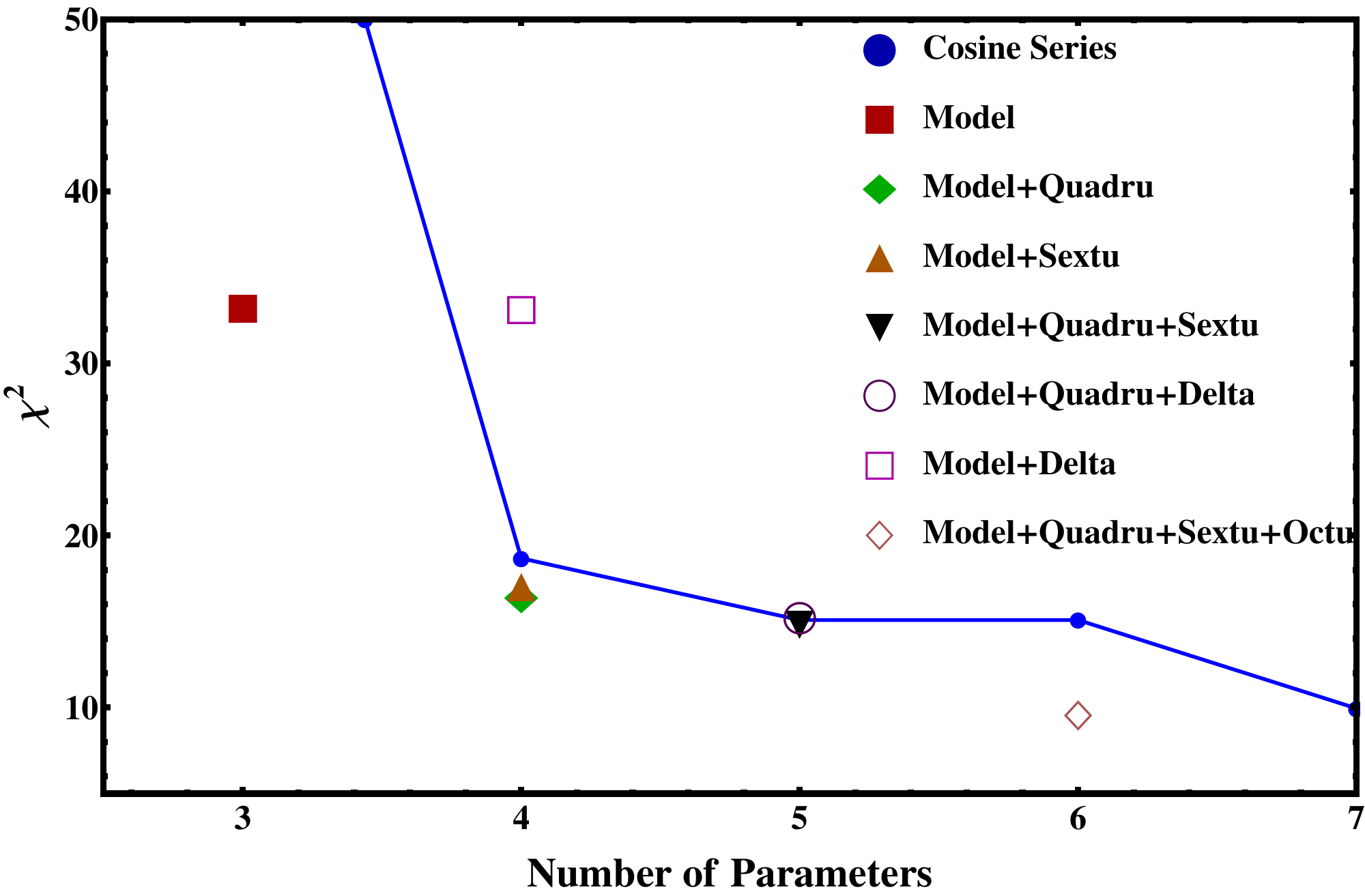}   
%\put(-28,90) {\bf (a)}
%\put(-158,120) {\bf bin 10}
\caption{\label{chi22} (Color online) $\chi^2$ values vs number of
  parameters $K$ for several data models. The general trend is again
  monotonic decrease with increasing parameter number.
} %AuAu200GeV-bin8-Chi2Modelsy
\end{figure}

Figure~\ref{chi22} shows $\chi^2$ values from conventional data
modeling. The FS-only model achieves a substantial reduction for $K =
4$ but no significant improvement for additional terms.  The basic
Model with $K = 3$ (solid red square) has a $\chi^2$ much elevated
from the number of fit DoF = 8 and is rejected on that basis. Addition
of a quadrupole component (solid green diamond) brings $\chi^2$ down
to an acceptable value. Addition of more cosines (sextupole, octupole)
to the basic Model + quadrupole tracks the FS-only noise
accommodation.

%%%%%%%%%%%%%%%%%%%%%%%
%%%%%%%%%%%%%%%%%%%%%%%

The basic Model + sextupole (solid red triangle) has the same $\chi^2$
value as that for basic Model + quadrupole. The Gaussian + dipole +
sextupole combination can interact to accommodate the independent
quadrupole component in the data, since the octupole component of the
Gaussian is only a few sigma above the statistical noise. Interactions
among the basic Model Gaussian and additional cosine terms are
discussed in Sec.~\ref{compete}.

%%%%%%%%%%%%%%%%%%%%%%%
\begin{figure}[h]  
\includegraphics[width=.48\textwidth]{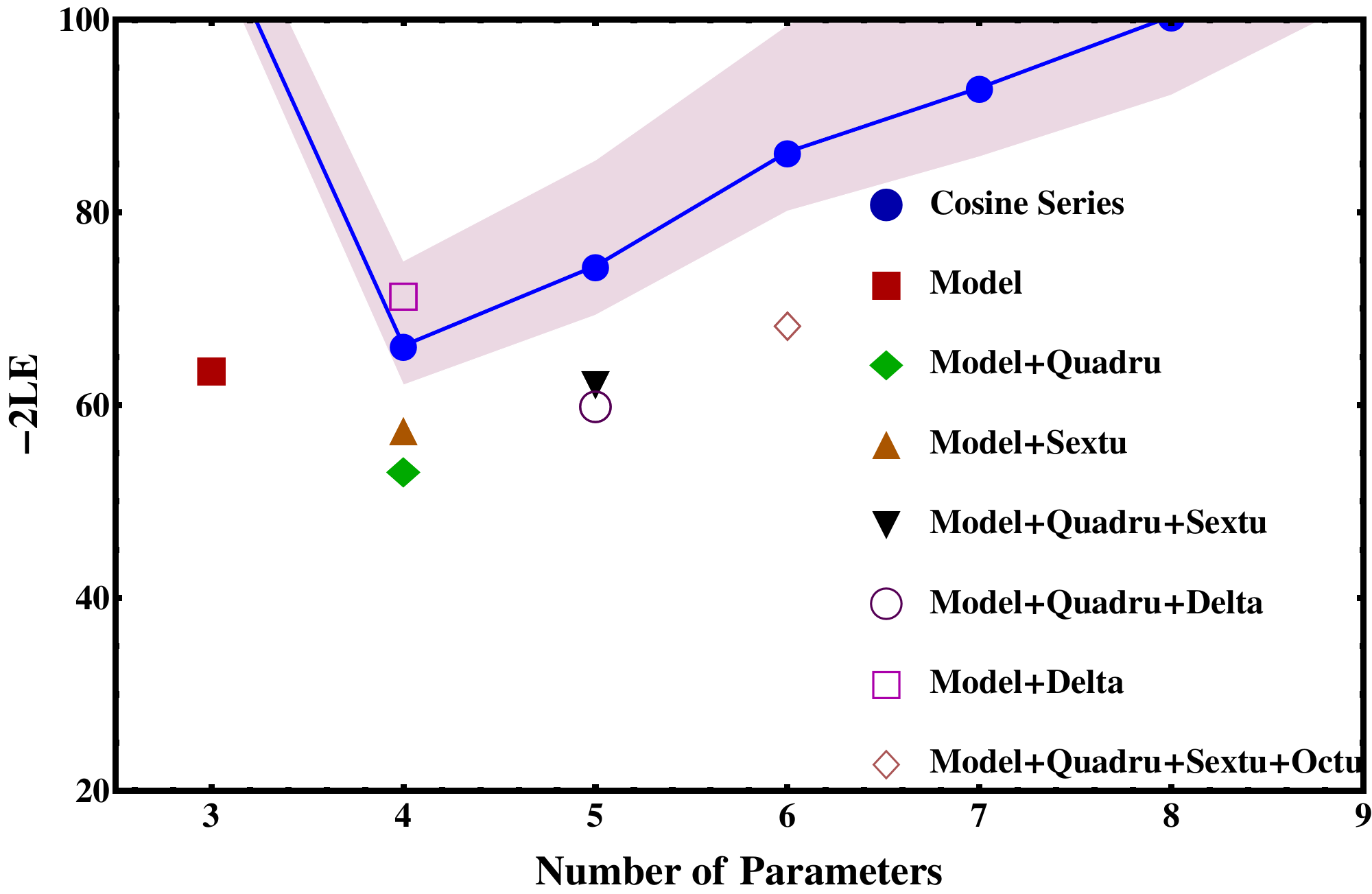}   
%\put(-28,90) {\bf (a)}
%\put(-158,120) {\bf bin 10}
\caption{\label{modelcomp2} (Color online) Negative log Evidence
  $-2LE$ vs number of parameters $K$ for several models. The basic
  Model + quadrupole (solid diamond) is strongly favored over others
  (lowest $-LE$, largest evidence). The basic Model alone (solid
  square) is strongly rejected by the evidence, as are FS-only models
  for all $K$ (blue points and line).
} %AuAu200GeV-bin8-EvidenceModels
\end{figure}
%%%%%%%%%%%%%%%%%%%%%%%

Figure~\ref{modelcomp2} shows negative log evidence $-2LE$ for various
models. The basic Model + quadrupole (solid green diamond)
corresponding to $K = 4$ model DoF has substantially smaller -2LE
(larger evidence $E$) than other model combinations. It is clearly
preferred over the basic Model alone by $\exp(5) \rightarrow 187 \pm
30$:1 odds. For other models the cost of extra parameters is not
justified by reductions in $\chi^2$. The quadrupole model component is
preferred over a sextupole by $\exp(2) \rightarrow 8.5 \pm 1.4$:1 due
to differences in the fit covariance matrix for the two models. All
FS-only models are again rejected by large factors.

%%%%%%%%%%%%
\section{bin-0 83-94\% Azimuth correlations} \label{bin0}

In this third of three cases, essentially representing \pp\ (\nn)
collisions, we encounter a major challenge for BI analysis from
several sources: (a) The SS peak on azimuth contains two contributions
that cannot be separated easily by discarding the bin at the origin as
they were for bins 8 and 10, (b) the signal amplitude is much smaller
relative to statistical noise (15:1) than it was for more-central
collisions (200:1), and (c) the SS peak is substantially broader on
azimuth.

\subsection{1D azimuth  projection}

\begin{figure}[h]  
\includegraphics[width=.48\textwidth]{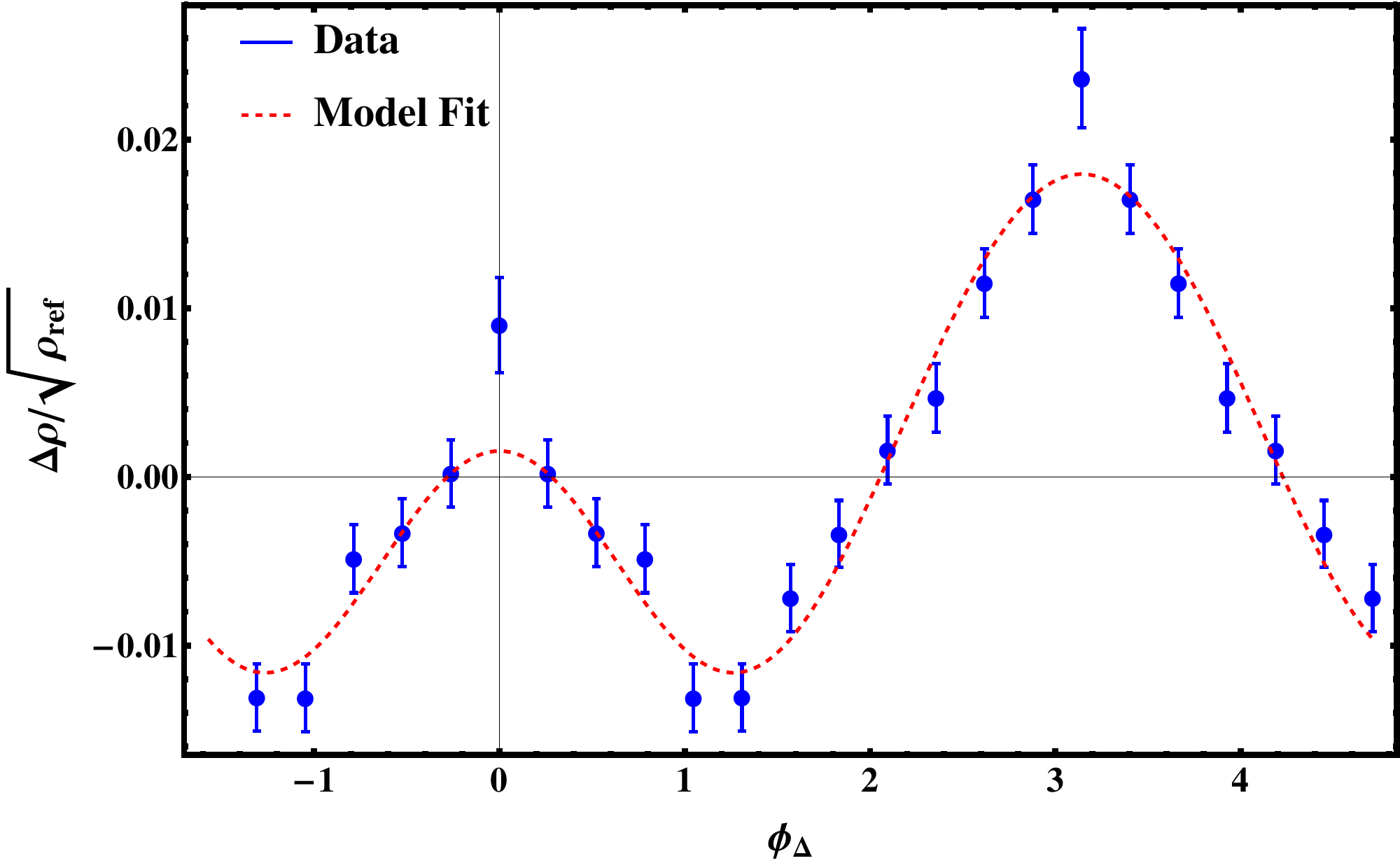}   
%\put(-28,90) {\bf (a)}
%\put(-158,120) {\bf bin 10}
\caption{\label{1ddata3} (Color online) 1D projection onto azimuth
  (points) from the 2D data histogram for 83 - 94\% central 200 GeV
  \auau\ collisions in Fig.~\ref{fig1} (a). The (red) dashed curve is
  a fit to the data with the basic Model of Eq.~(\ref{simpmod2}). The
  statistical errors are 0.0026.  } %AuAu2002GeV-bin0-Projection
\end{figure}

Figure~\ref{1ddata3} shows a projection of the 2D data histogram from
83-94\% central 200 GeV \auau\ collisions (points). Unlike previous
cases the SS peak includes a significant contribution from
BE/electrons that is not included in the models (conversion electron
pairs do fall mainly within the single bin at the origin).
A fit of the basic Model to data is shown by the dashed (red)
curve. The fitted model parameters are $A_{1D} =0.073\pm0.025$,
$\sigma_{\phi_\Delta} =0.926\pm0.128$ and $A'_D = 0.022\pm0.006$ with
$\chi^2\text{/DoF} = 12.0$ for 11 - 3 = 8 fit DoF.

%%%%%%%%%%%%%%%%%%%%%%%
%%%%%%%%%%%%%%%%%%%%%%%
    
\subsection{Data power spectrum} \label{powerspec2}

\begin{figure}[h]
%\put(-28,90) {\bf (a)}
\includegraphics[width=.48\textwidth]{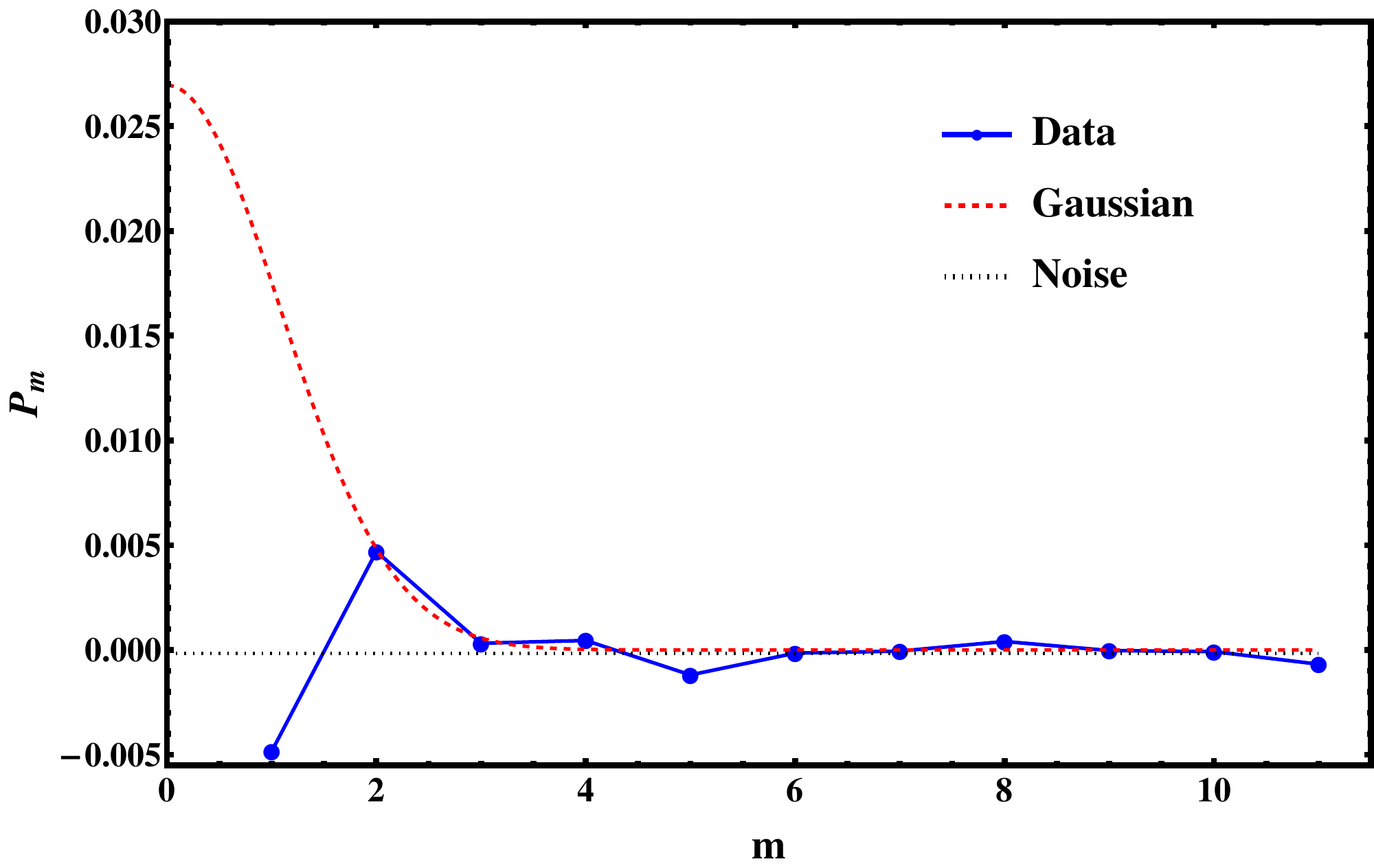}   
%\put(-158,120) {\bf bin 10}
\caption{\label{power3} (Color online) Power spectrum values $P_m$
  (points) derived from the data in Fig.~\ref{1ddata3} via
  Eq.~(\ref{coef}). The (red) dashed curve is the Gaussian PS
  described by Eq.~(\ref{fmm}) with Gaussian width and amplitude
  corresponding to the basic Model fitted to data in
  Fig.~\ref{1ddata3}. The interval $m > 2$ is consistent with a
  ``white-noise'' power spectrum (dotted line) representing the
  statistical noise in Fig.~\ref{1ddata3}.
} %AuAu200GeV-bin0-PowerSeries
\end{figure}

Figure~\ref{power3} shows the power spectrum (points and blue solid
curve) derived from the data in Fig.~\ref{1ddata3}.
As for previous centrality bins we include a predicted power spectrum
(red dashed curve) for a 1D Gaussian (SS peak) with amplitude and
width parameters derived from the basic Model fitted to data in
Fig.~\ref{1ddata3}.

%%%%%%%%%%%%%%%%%%%%%%%
%%%%%%%%%%%%%%%%%%%%%%%

Because the bin-0 SS peak is broader on azimuth (thus narrower on
index $m$) and the S/N is much smaller the PS signal is not
significant at $m = 4$ or even $m = 3$. A $K = 2$ FS-only model in the
form of dipole + quadrupole should be sufficient to displace the basic
Model. For bins 10 and 8 FS-only models are clearly excluded in favor
of the basic Model, but for bin 0 the $K=3$ basic Model and a $K=2$ FS
can both describe the two data DoF.  Thus, we expect BI analysis to
prefer the FS-only model.

\subsection{Bayesian model fits with Fourier series}

\begin{figure}[h]
\includegraphics[width=.48\textwidth]{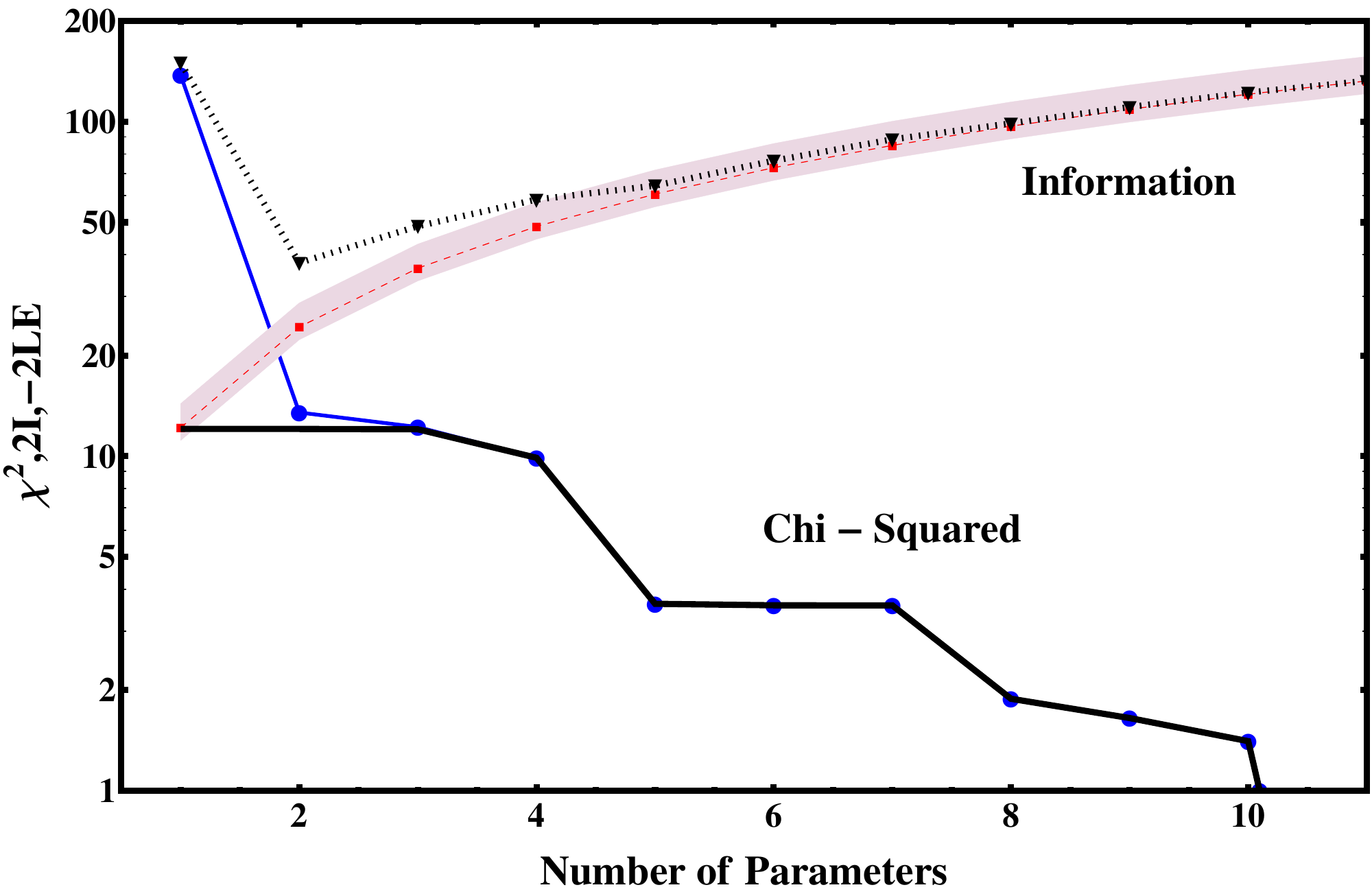}   
%\put(-28,90) {\bf (a)}
%\put(-158,120) {\bf bin 10}
\caption{\label{basicfit3} (Color online) $\chi^2$ (upper solid curve
  and points) and information $2I$ (dashed curve and points) vs number
  of parameters $K$ for FS-only models. The sum $-2LE$ (dotted curve
  and points) is also included. The $\chi^2$ trend for basic-Model fit
  residuals (lower solid curve) is approximately consistent with the
  expected noise trend $11 - K$.  } %AuAu200GeV-bin0-ChiSquared
\end{figure}

Figure~\ref{basicfit3} shows the FS-only $\chi^2$ trend for bin-0 data
(upper solid blue curve and points). As expected, $\chi^2$ drops to
the noise trend (lower solid curve) by $K = 2$. Additional terms
accommodate statistical noise. Information $2I$ follows the expected
monotonic increase $\approx 10K$.  Negative log evidence $-2LE$ has a
minimum for $K = 2$. Thus, an FS-only model with $K = 2$ is preferred
by the data, as expected from the PS in Fig.~\ref{power3}.

%%%%%%%%%%%%%%%%%%%%%%%
%%%%%%%%%%%%%%%%%%%%%%%

\subsection{Bayesian model comparisons}

\begin{figure}[h]  
\includegraphics[width=.48\textwidth]{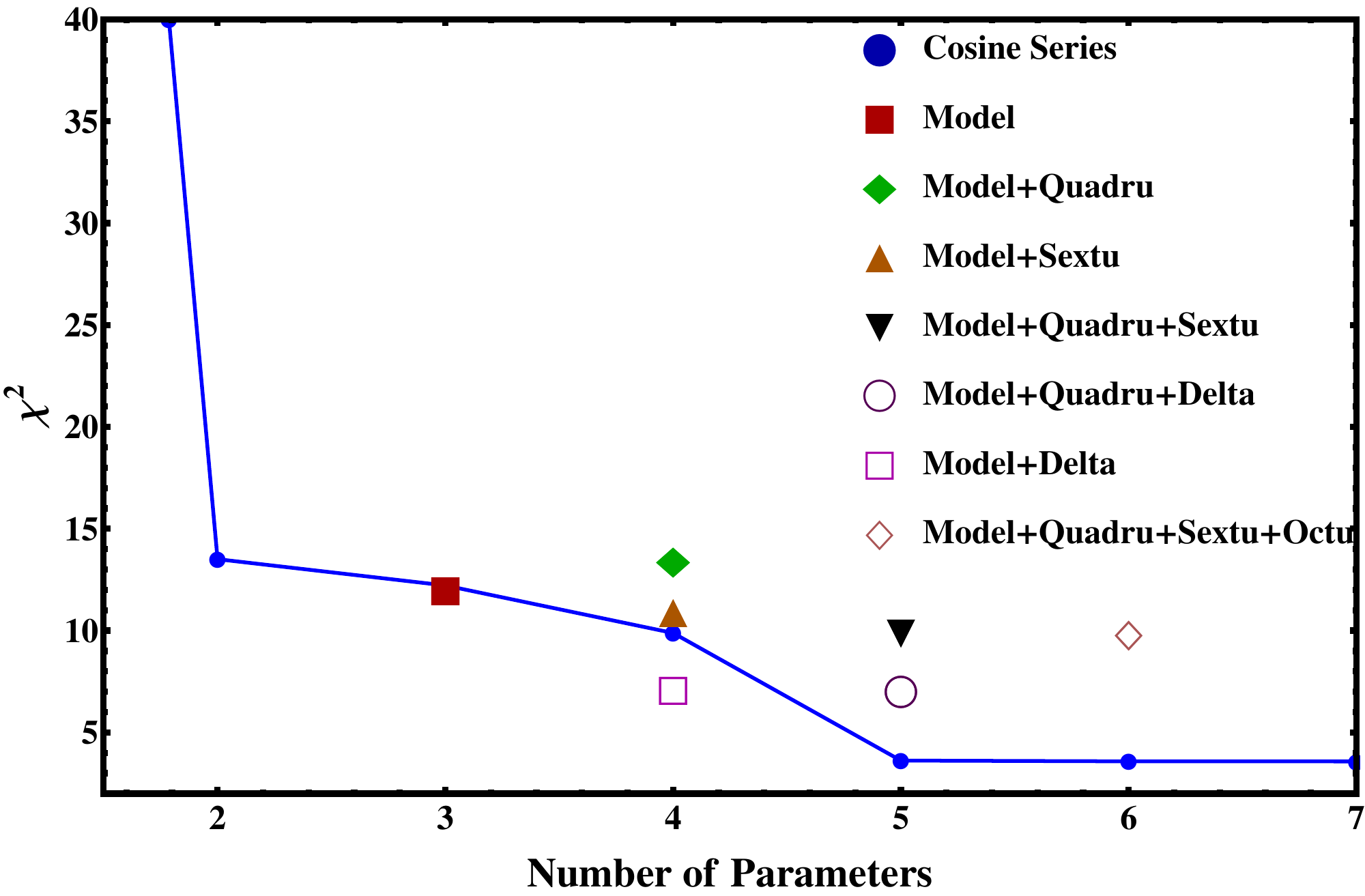}   
% \put(-28,90) {\bf (a)} \put(-158,120) {\bf bin 10}
\caption{\label{chi23} (Color online) $\chi^2$ values vs number of
  parameters $K$ for several data models. The basic Model (solid
  square) is equivalent to the FS-only model with $K = 3$ (solid dot).
} %AuAu200GeV-bin10-Chi2Modelsy
\end{figure}

Figure~\ref{chi23} shows $\chi^2$ trends for several competing models
applied to the bin-0 data in Fig.~\ref{1ddata3}. The $K=3$ basic Model
and FS-only model describe the data equally well, and we expect the
simpler $K = 2$ FS-only model to be preferred when an Occam penalty is
included. For these bin-0 data the addition of a ``delta'' component
at $\pi$ (open square) leads to substantial improvement in the fit
quality, consistent with Fig.~\ref{1ddata3}.

%%%%%%%%%%%%%%%%%%%%%%%
%%%%%%%%%%%%%%%%%%%%%%%

\begin{figure}[h]  
\includegraphics[width=.48\textwidth]{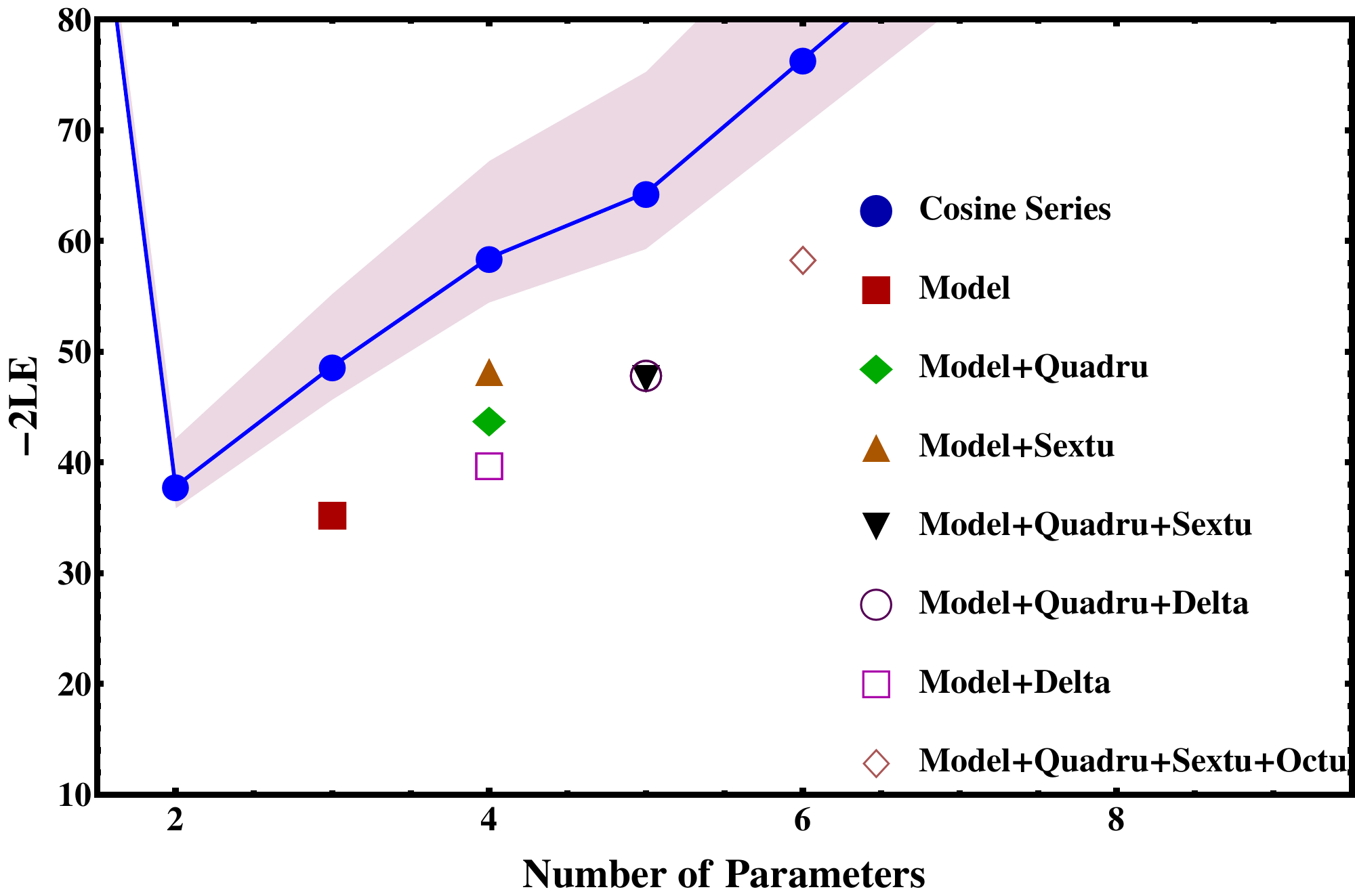}   
%\put(-28,90) {\bf (a)}
%\put(-158,120) {\bf bin 10}
\caption{\label{modelcomp3} (Color online) Negative log Evidence
  $-2LE$ vs number of parameters $K$ for several models.  The $K = 3$
  basic Model (solid square) is favored over all others (lowest $-LE$,
  largest evidence), especially over the $K = 2$ FS-only model
  expected to prevail for this centrality (lowest solid dot).
} %AuAu200GeV-bin10-EvidenceModels
\end{figure}

Figure~\ref{modelcomp3} shows the log evidence $-2LE$ trend. That the
$K = 3$ basic Model (solid red square) is preferred over the $K=2$
FS-only model (lowest blue point) despite the cost of the extra model
parameter is a major surprise. The evidence ratio (odds) is $3.3 \pm
0.25$:1 $ \approx \exp(1.25)$. That result prompted a detailed study
reported in Sec.~\ref{why} on how information $I$ is related to model
priors and data, with supporting material provided in
App.~\ref{structure}.

%%%%%%%%%%%%%%%%%%%%%%%
%%%%%%%%%%%%%%%%%%%%%%%

The ability in this case to discriminate between the basic Model and
FS models, despite 1D data with low S/N ratio, is a significant
achievement for BI analysis.
The correctness of the basic-Model preference is confirmed by analysis
of 2D data histograms. From the 2D analysis of Ref.~\cite{anomalous}
we learn that the SS 2D peak is {\em necessary} for all
centralities. In contrast, a 1D FS-only model would fail dramatically
for any 2D data, but that is not apparent from 1D projections alone.

%%%%%%%%%%%%%%%%
\section{Systematic uncertainties} \label{system}

Bayesian Inference methods provide a powerful system for
discriminating among competing complex data models with a consistent
set of evaluation rules. Close examination of method details and
evaluation of uncertainties is required to insure confidence in the
results.

\subsection{Uncertainties for data histograms}

For 2D histograms from Ref.~\cite{anomalous} the angular acceptance
was divided into 25 bins on the $\eta_{\Delta}$ axis and 25 bins on
$\phi_{\Delta}$, a trade off between statistical error magnitude and
angular resolution. The histograms are by construction symmetric about
$\eta_\Delta = 0$ and $\phi_\Delta = 0, \pi$. The 25 bins on
$\phi_\Delta$ actually span $2\pi + \pi/12$ to insure centering of
major peaks on azimuth bin centers. 2D binwise statistical errors are
$\pm 0.004$ for 200~GeV data near $|\eta_{\Delta}| = 0$. Because of
the $\eta_\Delta$ dependence of the pair acceptance statistical errors
increase with $|\eta_\Delta|$ as $\sqrt{\Delta\eta/(\Delta\eta -
  |\eta_{\Delta}|)}$ with $\eta$ acceptance $\Delta\eta = 2$.
Errors are uniform on $\phi_{\Delta}$ except that errors are larger by
factor $\sqrt{2}$ for angle bins with $\phi_{\Delta} = 0$ and $\pm
\pi$ because of reflection symmetries.

Statistical errors are approximately independent of centrality for the
per-particle statistical measure $\Delta \rho /
\sqrt{\rho_\text{ref}}$ over nine 10\% centrality bins (0-8).  An
additional factor $\sqrt{2}$ increase applies to the two most-central
centrality bins (9, 10) which split the top 10\% of the total cross
section. After projection onto 1D azimuth for this study the
centrality bin 10 errors are about 0.0037 except for the azimuth bins
at 0 and $\pi$. Errors for the other centrality bins (0, 8) are a
factor $1/\sqrt{2}$ less or 0.0026. $\chi^2$ values for optimized
models in this study determined with those statistical errors are
generally consistent with the number of fit DoF = data DoF $-$ K
(number of model parameters), as demonstrated in
Fig.~\ref{basicfit}. Thus, the statistical and systematic
uncertainties for data histograms used in this study are both small
and well understood.

\subsection{Uncertainties for information estimation} \label{infosys}

Information uncertainty is largely related to the choice of prior PDFs
for various model parameters and the fitted-parameter
uncertainties. We repeat the information definition in
Eq.~(\ref{infoeq1})
\begin{eqnarray} \label{infoeq}
I(D^*|H) &=& -\ln\left[ \sqrt{(2\pi)^K \det C_K}~ p(\tilde w|H) \right],
\end{eqnarray}
the natural log of prior volume $V_w(H)$ over posterior volume
$V_w(D^*|H)$ in the model parameter space.  The covariance matrix
$C_K$ for a $K$-parameter model is obtained from the Hessian
describing the curvatures of the likelihood function near its
maximum. The likelihood function is in turn determined by model $H$ in
combination with specific data $D^*$. If the model and prior are
defined and data specified the posterior volume is also well defined.

Assuming translation invariance within a parameter-space volume where
the likelihood is significantly nonzero the prior PDF for parameter
$w_k$ is taken to be uniform across a bounded interval $\Delta_k$ into
which the corresponding fitted parameter value should almost certainly
fall. The prior volume for $K$ parameters is then
\begin{eqnarray} 
  1/p(\tilde w|H) &=& \prod_{k = 1}^K \Delta_k \equiv V_w(H). %
\end{eqnarray} %
In principle, a prior is defined before data $D^*$ are obtained and
thus should not depend on specific data. However, it is fair to invoke
general knowledge ($Q$) about the typical amplitudes of structures in
such data. We know from experience that typical structure amplitudes
(e.g.\ peak-to-peak excursions) are generally $< O(1)$. That applies
for example to the Gaussian amplitude in the basic Model, and to the
Gaussian width based on the definition of the SS peak, implying that
$\Delta_k \approx 1$ in those cases.

What matters more than absolute estimates of $\Delta_{k}$ is the
relations among different models and model parameters. If
prior-interval estimates are excessive for a particular model it may
be unduly penalized. Given the above assignment for a Gaussian
amplitude, what is a fair assignment for cosine coefficients? To that
end we examine Eq.~(\ref{wiener1}). The autocorrelation to be modeled
on the left receives contributions at the origin from several FS
components $P_m$ including factors 2. Thus, it is reasonable to assume
that cosine coefficients and uncertainties may be substantially
smaller on average than the Gaussian amplitude and uncertainty. For
all cosine components we assign $\Delta_{k} = 1/3$ and indicate
prior-related uncertainties by including $\Delta_{k} = 1$ and
$\Delta_{k} = 1/5$ as limiting cases for cosines (e.g.\ curve $I$ and
hatched band in Fig.~\ref{basicfit}).

If we assume equal prior intervals $\Delta_k$ and equal variances
$\sigma_k^2$ for $K$ model parameters and negligible covariances among
parameters information $I$ simplifies to
\begin{eqnarray}
  I &\approx& K \left\langle \ln \left[ \frac{\Delta_k}{\sqrt{2\pi}\sigma_k} \right] \right\rangle + \text{constant}.
\end{eqnarray}
In fits with FS-only models we observe $\sigma_k \approx
O(0.0007)$. Given $\Delta_k \approx 1$ we have $\ln(\Delta_k / \sqrt{2
  \pi} \sigma_k) \approx 6$, while if we reduce to $\Delta_k
\rightarrow 1/3$ (assumed for all cosine terms) $\ln(\Delta_k /
\sqrt{2 \pi} \sigma_k) \approx 5$. If we further reduce $\Delta_k
\rightarrow 1/5$ with $\ln(\Delta_k / \sqrt{2 \pi} \sigma_k) \approx
4.7$ the fitted parameter values in some cases contradict the prior,
implying that the chosen prior interval is too small.
We can then state that for all FS-only models $I/K = 5.3 \pm 0.6$. For
the basic model with added cosines the parameter uncertainties are
more typically $\sigma_k \approx O(0.01)$. In that case we obtain $I/K
= 2.6 \pm 0.5$. Those results imply that addition of a model parameter
is justified ($-2LE = \chi^2 + 2I$ is significantly reduced) if the
resulting decrease in $\chi^2$ is significantly greater than $2I/K
\approx 10$ for FS-only models and $\approx 5$ for the basic Model
plus optional cosines.

\subsection{ Uncertainties for odds ratios} \label{oddsys}

As noted in Sec.~\ref{modcompare} odds ratios can be used to state
quantitatively the BI relation between two models in the form of a
probability ratio $p(D^*|H_1) / p(D^*|H_2) \rightarrow E(D^*|H_1) /
E(D^*|H_2)$, where equality to the second ratio assumes equal {\em
  model} priors $p(H)$ for the two cases. In terms of log evidence
$LE$ the Bayes Factor is $B_{12} = \ln[E(D^*|H_1) / E(D^*|H_2)]$, and
the odds is then $\exp(B_{12})$.

The uncertainty (error) in an odds ratio is determined by the
uncertainties in the compared evidences in turn dominated by
uncertainties in the covariance matrix/Hessian and the prior
PDFs. Uncertainties for the Hessian matrix are discussed in
App. \ref{hesserror} and serve as the sole basis for the odds errors
stated in the text.

Uncertainties for the prior PDFs are discussed in the previous
subsection. The priors for SS Gaussian amplitude and width are set to
the minimum values consistent with experience, {\em disfavoring the
  basic Model a priori} and implying that any odds favoring the basic
Model is a lower limit.  A common uncertainty of a factor 2 either way
is assumed for a cosine coefficient in any model. Because an odds
estimate is a probability ratio systematic errors correlated between
numerator and denominator cancel in first order, whereas uncorrelated
random errors should combine quadratically. That property can be seen
as an advantage for odds as a basis for model comparisons and
minimizes the uncertainty contribution from cosine elements common to
two compared models.

If models with different $K$ values (parameter number) are compared
the unpaired systematic error is not canceled. For instance, the odds
between the basic Model ($K = 3$) vs FS-only model ($K=4$) for bin 10
includes a linear dependence on the FS-only prior uncertainty for one
additional cosine term. However, the comparison of basic Model plus
quadrupole vs FS-only model for bin 8 (both $K = 4$) eliminates that
uncertainty contribution.

%%%%%%%%%%%%%%%%%%%%%%%%%%%%%%%
\section{Discussion} \label{disc}

We consider several issues that have arisen in application of BI
methods to azimuth-correlation data models, including surprising
performance of the 1D basic Model in peripheral collisions, consistent
strong preference for the basic Model by BI analysis, competition
between Gaussian and cosine terms in data models, and implications
from this study for two theoretical narratives.

\subsection{Comparing bin 10 and bin 0}

The data structure for centrality bin 10 in Fig.~\ref{1ddata} could be
modeled as (a) two peaks at 0 and $\pi$, (b) as a Fourier series only,
or (c) as a combination of such elements. The two peaks described by
the basic Model are expected in a HEP/jets narrative describing
high-energy nuclear collisions. FS models are expected in a QGP/flow
narrative and are capable of describing any structure on periodic
azimuth. Competition among data models thus reflects competition
between theoretical narratives.

In Fig.~\ref{power1} we learn that all information in the data PS is
confined to $m \in [1,4]$. Higher terms in an FS model describe only
statistical noise. Comparing a PS Gaussian model with the data PS we
find that four points are predicted by a Gaussian fitted to data, and
one point corresponds to the fitted dipole within the basic Model. The
$K = 3$ basic Model fully represents the data signal as demonstrated
in Fig.~\ref{noise}, but so does a $K = 4$ FS-only model. Intermediate
combinations of Gaussian + cosines also describe the data well. Models
with more parameters continue to reduce $\chi^2$ as in
Fig.~\ref{chi2}, and might be preferred on that basis.

However, when an Occam penalty is introduced in the form of
information $I$ dramatic differences among models appear, as in
Figs.~\ref{modelcomp} and \ref{xxxx}. In the latter figure the
basic-Model probability is $p(H_l) \approx 80$\%, the next highest
being basic Model + quadrupole with $p(H_l) \approx 15$\%. Adding more
cosine components may reduce $\chi^2$, but not to an extent that
compensates large Occam penalties (increase in $I$). The additions are
essentially ``fitting the noise'' and are strongly rejected by BI
analysis.

A different situation emerges for Bin 0. In Fig.~\ref{power3} the data
signal is confined to $m \in [1.2]$ for two reasons: (a) The S/N ratio
is reduced by a factor 13 and (b) the SS peak azimuth width is
increased by 30\% so the conjugate PS signal peak width is reduced by
that factor. Consequently the ``bandwidth'' of the data PS signal is
reduced from $m \in [1,4]$ to $m \in [1,2]$. A $K=2$ FS-only model
with two parameters should then be strongly preferred by BI analysis
over the $K=3$ basic Model, given equivalent priors for the two
models.

However, that is not what we find in Fig.~\ref{modelcomp3}. The basic
Model maintains a significant advantage over a $K=2$ FS-only model,
the odds ratio being $\approx \exp(1.25) = 3.5:1$ in favor of the
basic Model, even with one more parameter. That surprising result led
to the detailed comparisons in the next subsection and the study in
App.~\ref{structure}.

The 1D projection is not the only information we have about the source
of these data. The unprojected 2D histogram in Fig.~\ref{fig1} (a)
clearly indicates that a SS 2D peak model is required by bin-0 data,
and a 2D FS-only model would be rejected by a large
factor~\cite{anomalous,multipoles}. The 2D observations contribute a
larger Bayesian context ($Q$) applicable to this 1D BI study. The
FS-only model is ruled out for all 2D histograms as shown in previous
studies~\cite{multipoles}.

\subsection{Why BI analysis favors the basic Model} \label{why}

The basic Model alone (for near-central collisions) or the basic Model
plus quadrupole (for noncentral collisions) is strongly preferred by
BI analysis over FS-only models, even for the most-peripheral
collisions where the 1D data include only two significant DoF. The
choice of priors is not the reason; priors are applied consistently
for each parameter type within any model. The large difference in
evidence values is dominated by differences in the fit covariance
matrix. The r.m.s.\ parameter errors for FS models are consistently
10-15 times smaller than for the basic Model. Evidence differences
correspond to differences in model {\em predictivity}, as illustrated
in the following comparison and App.~\ref{structure}.

\begin{figure}[h]  
%\put(-28,90) {\bf (a)}
\includegraphics[width=.48\textwidth]{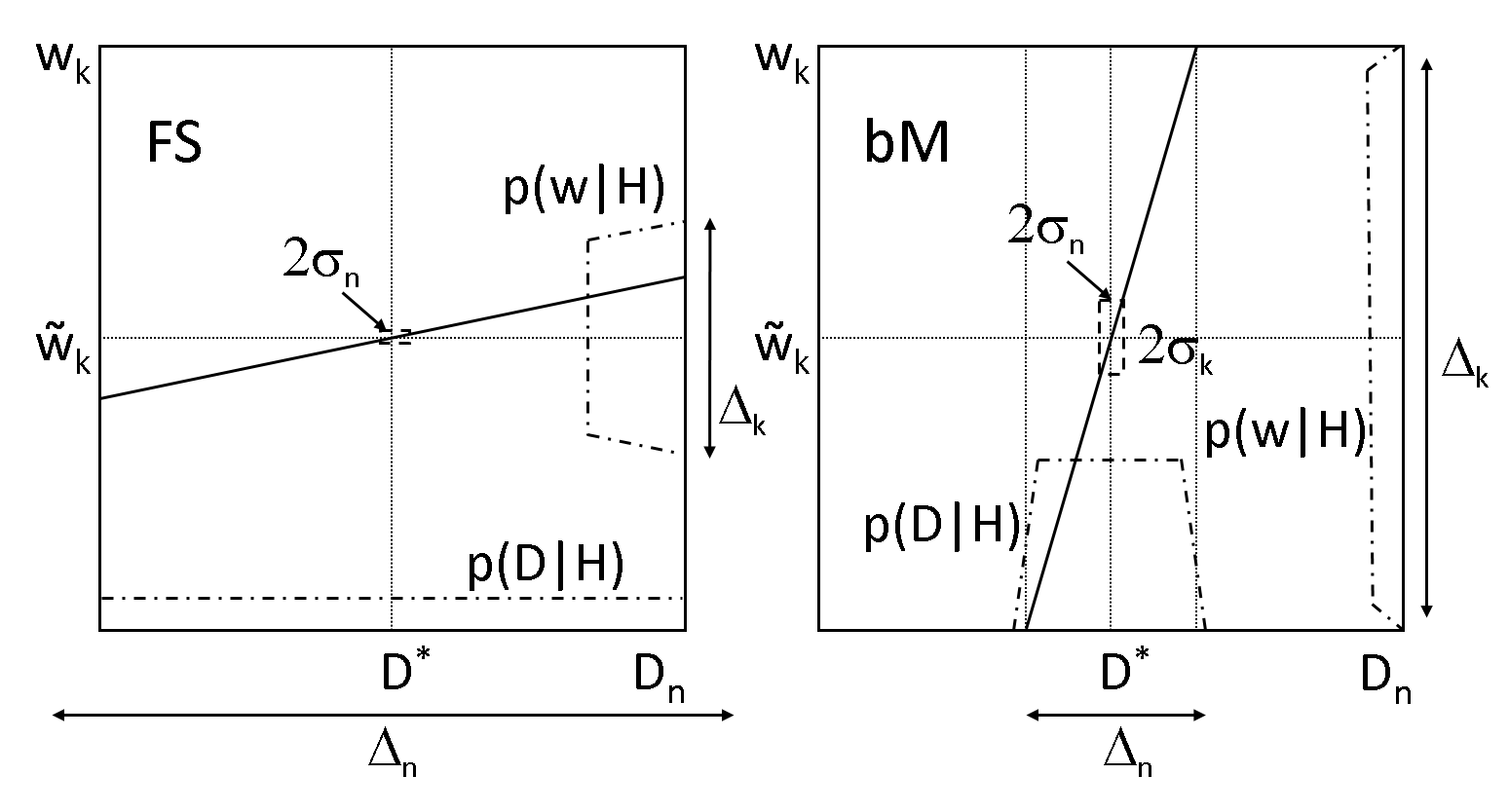}   
% \put(-158,120) {\bf bin 10}
\caption{\label{pairpanel} Joint parameter-data space for two data
  models.  These panels are zoomed out from the scale of
  Fig.~\ref{bayessp} to reveal the PDFs distributed over the entire
  parameter and data spaces. Given prior intervals $\Delta_k$ the
  model angles $\theta_{kn}$ determine the magnitudes of the predicted
  data intervals $\Delta_n$, the predicted data volume $V_D(H)$, and
  therefore the evidence $E( D^*|H)$. The $\theta_{kn}$, and hence the
  Jacobian of a model function, largely determine the predictivity of
  a model. For these two models the typical $\tan(\theta_{kn})$ differ
  by factor 12.  } %pair-panel
\end{figure}

Figure~\ref{pairpanel} shows sketches of joint data-parameter spaces
for an FS-only model (left) and the basic Model (right) corresponding
to bin-10 data. The parameter errors for the FS-only model are
typically $\sigma_k \approx 0.0007$. The parameter errors for the
basic Model in Table~\ref{paramx} are 0.007 for SS peak amplitude and
width and 0.002 for dipole amplitude, but with added cosine terms the
errors increase to $\approx 0.01$. The data errors for bin 10 are
$\sigma_D \approx 0.0037$. In terms of angles $\theta_{kn}$ defined in
Eq.~(\ref{angles}) $\tan(\theta_{kn}) \approx$ 1/3 for the basic Model
(20\degree) and 5 for the FS-only model (80\degree). The errors and
$\tan(\theta_{kn})$ are represented by the dashed rectangles and the
angles of the diagonals (solid lines) in the two panels, as in
Fig.~\ref{bayessp}.

%%%%%%%%%%%%%%%%%%%%%%%
%%%%%%%%%%%%%%%%%%%%%%%

In Fig.~\ref{pairpanel} the prior PDFs are represented by the vertical
dash-dotted lines in each panel and the arrows labeled $\Delta_k$. For
all cosine amplitudes the prior is $\Delta_k \approx 1/3$. For the SS
peak amplitude and width the priors are $\Delta_k \approx 1$. We can
estimate the evidences or predicted data volumes based on the argument
in App.~\ref{structure} where the relation between data-space volume
and parameter-space volume is determined by angle factors
$\tan(\theta_{kn})$. For the basic Model (right panel) even the larger
priors ($\Delta_k \approx 1$) are mapped to smaller data intervals
($\Delta_n \approx 1/3$), whereas for FS-only models (left panel)
smaller priors ($\Delta_k \approx 1/3$) are mapped to larger data
intervals ($\Delta_n \approx 5/3$).
The result is {\em much smaller predicted data volumes} $V_D(H)$ for
the basic Model, and consequently much higher evidence and
plausibility compared to FS-only models.

The same argument applies to changes in evidence or information with
increasing parameter number (e.g.\ added cosine terms). From
Eq.~(\ref{edh}) (assuming comparable $\chi^2$ values for competing
models)
\begin{eqnarray} \label{iconst} I &=& \sum_{k=1}^K \ln
  \left[\frac{\Delta_k}{\sqrt{2\pi}\sigma_k} \right] +
  \text{constant}.
\end{eqnarray}
With $\Delta_k \approx 1/3$ for all cosine terms, $\sigma_k \approx
0.0007$ for FS-only models and $\approx 0.01$ for basic Model +
cosines the typical increment per cosine term is $I/K \approx 5$ for
FS-only models and 2.5 for basic Model + cosines, e.g.\ consistent
with Fig.~\ref{modelcomp}. Thus, evidence and information trends are
determined mainly by the relative parameter errors reflecting the
Jacobians and model algebra.

The fitted-parameter errors reflect the algebraic structure of the
model, as discussed in App. A. Because a Fourier series is orthogonal
each coefficient is determined independently. Since the Fourier model
elements individually do not resemble the data there is required a
very "fragile" assembly of terms that easily overfits the data (treats
noise as signal). Only a small range of FS-only parameter values can
reproduce a given data set, and the parameter variances are
consequently very small.

In contrast, the basis Model includes a Gaussian (motivated by the
data structure) with nonlinear parameter $\sigma_{\phi}$ that covaries
with other parameters. Thus, larger ranges of basic-Model parameters
can reproduce the data adequately, and the parameter variances are
correspondingly larger. The basic Model is more "robust" because on
average the model elements {\em individually} look more like isolated
data components. If both models give the same chi-squared fit the
basic Model is preferred by BI analysis because on average it is far
more likely to describe the data accurately (a larger fraction of the
prior-delimited parameter space provides an acceptable description for
the given data).

We conclude that the key issue for Bayesian model comparisons is model
predictivity.  The basic Model is highly predictive (therefore
falsifiable), describing two peaks (fixed at 0 and $\pi$), with one
peak as wide as possible and the other somewhat narrower. Two peak
amplitudes and a width are the only parameters. The basic Model is
consistent with the HEP/jets narrative but was inferred from data
without any theory assumptions. In contrast, FS-only models can
describe any structure on azimuth, have no predictivity (are not
falsifiable) and are therefore strongly rejected by BI analysis. Model
predictivity [smallness of predicted data volume $V_D(H)$] is
determined largely by the algebraic structure of the data model
(Jacobian) as revealed by fitted-parameter errors compared to data
errors via the $\tan(\theta_{kn})$ elements.

%%%%%%%%%%%%%%%%%%%%%%

\subsection{Competition: extra cosine terms vs SS Gaussian} \label{compete}

The bin-10 results in Table~\ref{paramx} can be used to examine the
consequences of adding one or more cosine terms to the basic Model
when there is no corresponding data signal. The $\chi^2$ is reduced in
general, suggesting an improved data description. However, in some
cases the model parameters undergo large changes seeming to indicate
that model parameters are very uncertain.
To understand the apparent contradiction we consider the bin-10
``worst case'' model (basic Model + quadrupole + sextupole + octupole)
appearing in the next-to-last column of Table~\ref{paramx}.

The model difference (``$+ A_O$'' $-$ basic Model) for each cosine
coefficient is $\Delta A'_D = 0.115$, $\Delta A_Q = 0.064$, $\Delta
A_S = 0.024$ and $\Delta A_O = 0.005$ for $m \in [1,4]$. The
differences correspond to the predicted Gaussian PS values in
Fig.~\ref{power1} (red dashed curve). In effect, changes in the cosine
coefficients of the FS-only model are equivalent to the fitted
Gaussian already describing the data signal correctly in the basic
Model. The SS Gaussian required by the data is effectively excluded
from the data model by the added cosine terms, reduced to a minor
role~\cite{sextupole}.

The bin-10 result reveals a competition between the basic Model and a
truncated FS to describe signal + noise. The competing truncated FS
offers more flexibility in accommodating noise compared to the
monolithic Gaussian. The FS may ``win'' in terms of $\chi^2$, but a
well-chosen model element (Gaussian) describes only the signal and
excludes the noise.
Referring to Fig.~\ref{chi2} the $K = 6$ (``+ $A_O$'') model (open
diamond) has the same $\chi^2$ value as the $K = 4$ FS-only model
(solid point) because the former is effectively a $K = 4$ FS. The
Gaussian, with two parameters, has been excluded from the fit model
owing to noise competition, but its two parameters still contribute to
the Occam penalty. BI analysis then rejects the unnecessary cosine
terms in favor of the basic Model.

\subsection{Evidence extending beyond single histograms}

A model may describe data from some \aa\ centralities well but others
poorly. Nevertheless, the model may be retained by convention because
of desirable features (such as flow interpretations). Other forms of
data selection (\pt\ cuts, 1D projections, ratio measures) present
similar issues. In response we propose to extend BI methods beyond
single data histograms, combining results into one comprehensive
evaluation for competing data models.

The mechanism is suggested by the nature of Bayesian evidence $E$. The
evidence is a {\em probability}, and by the rules governing
probabilities the {\em joint} evidence for several cases should be the
{\em product of elementary evidences} (assuming approximate
independence).  For instance, the evidence for a model of 200 GeV
\auau\ collisions should be the product of evidences for individual
centralities. If a model claiming to describe all data components is
falsified for one component then it is falsified for all.
More generally, a model that provides an adequate description for all
cases may be preferred over a model that is favored for some cases but
strongly disfavored for others.

That principle extends not only to \aa\ centralities but to different
collision energies, \ab\ collision systems, spectrum and correlation
measures and data cuts. Evidence $E$ as a {\em product measure}
introduces an ``and'' condition for data description. A candidate
model {\em must} address all available data within its parameter space
or be rejected.

\subsection{Implications for theoretical narratives}

As noted in the introduction HEP/jets and QGP/flow narratives
currently compete to describe and interpret high-energy
nuclear-collision data through choices of data model and emphasis on
specific data and measured quantities. The HEP/jets narrative predicts
two dijet-related peaks on 1D azimuth, just what the basic Model
describes. Almost all 1D azimuth correlation data from the RHIC are
described by the basic Model + quadrupole with modest parameter
variations.
The QGP/flow narrative prefers various forms of the FS-only data model
interpreted physically in a flow context, from a single cosine ($v_2$,
index $k =2$) to several cosines interpreted to include ``higher
harmonic'' flows (index $k \in [1,5]$).

In the present study we apply BI methods to 1D azimuth data models
associated with the two narratives. BI analysis strongly favors the
basic Model in all cases, combined with an additional quadrupole
$\cos(2\phi)$ term except for the most-central data. The FS-only model
is strongly rejected in all cases. As discussed in Sec.~\ref{why} and
App.~\ref{structure} the main reason for BI rejection is lack of
predictivity for FS-only models, whereas the basic Model is strongly
predictive and therefore falsifiable. The present BI analysis thus
seems to support the HEP/jets narrative and reject the QGP/flow
narrative per their data models.

It could be argued that application of BI methods to data models
represents an arbitrary choice motivated by interest in a specific
outcome. However, we are faced with the requirement to evaluate
conflicting data models according to some neutral criteria. $\chi^2$
minimization always prefers more-complex data models that may reveal
little about data structure and possible physical mechanisms. Flow
interpretations are always possible for FS-only models, but such
models {\em cannot exclude a dijet interpretation} since they are able
to describe any data configuration.

Additional criteria are therefore required to test data
models. Guidance as to choice is provided by the role of rational
inference within the scientific method. It is recognized that physical
theories cannot be {\em proven}, can only be {\em falsified} by data,
requiring that candidate theories be {\em predictive}. Unpredictive
theories are not falsifiable and are therefore rejected as
candidates. In a Bayesian context predictivity is measured by
information $I$ and evidence $E$ as demonstrated in this study.
For a well-tested physical theory $H$ encountering new data $D^*$ the
information $I \approx 0$, and the predicted data-space volume
$V_D(H)$ is small.
If $D^* \notin V_D(H)$ the theory is falsified but $D^* \in V_D(H)$
results in plausibility $p(H|D^*) \to 1$: dramatically different
results

In the present analysis we encounter not competing physical theories
but competing data models serving as proxies. BI analysis evaluates
data models according to predictivity, i.e.\ the degree of restriction
on allowed data configurations. We conclude that the basic Model with
optional quadrupole component is very predictive, corresponding to
small information gain from newly-received data and consequent small
predicted data-space volume. FS-only models are not predictive, can
accommodate any data configuration, and are therefore rejected.

%%%%%%%%%%%%%%%%%%%%%%%%
\section{Summary and Conclusions} \label{summ}

Based on data from the relativistic heavy ion collider (RHIC) and
large hadron collider (LHC) claims have been made for formation in
high-energy nucleus-nucleus (\aa) collisions of a strongly-coupled
quark-gluon plasma (sQGP) with small viscosity -- a ``perfect
liquid.'' Such claims are based mainly on measurements of Fourier
coefficients $v_m$ of cosine terms $\cos(m\phi)$ used to describe
two-particle correlations on azimuth $\phi$ and interpreted to
represent flows, especially $v_2$ representing elliptic flow. In the
flow context dijets play a comparatively negligible role in
final-state correlation structure.

Modeling azimuth correlations by truncated Fourier series or
individual cosine terms is not unique. Other model functions can
describe the same data equally well and do suggest alternative
physical interpretations, especially substantial contributions from
dijet production. In effect, two physics narratives compete to
describe and interpret the same data. In one narrative collision
dynamics is dominated by dijet production. In the other narrative
collision dynamics is dominated by a dense, flowing QCD medium.
Opposing narratives appear to be supported by their respective data
models. To break the deadlock a method is required to evaluate model
functions according to neutral criteria and identify a preferred
model.

In this study we introduce {\em Bayesian Inference} (BI) to evaluate
competing model functions. BI analysis relies on a combination of the
usual $\chi^2$ goodness-of-fit parameter and {\em information} $I$
derived from the fit covariance matrix. $I$ quantifies changes in the
data model arising from acquisition of new data and represents an {\em
  Occam penalty} for excessive model complexity. Combination $\chi^2/2
+I$ leads to {\em evidence} parameter $E$ that determines the {\em
  plausibility} of each model when confronted with new data values.
The goal is to rank data models according to BI criteria without
resorting to {\em a priori} physics assumptions.

We apply several representative model functions to angular correlation
data and evaluate the model performance with BI methods. The data are
published 2D angular correlations from three centrality classes of 200
GeV \auau\ collisions on $(\eta,\phi)$. 2D histograms are projected
onto periodic azimuth $\phi$ by integration over pseudorapidity
$\eta$.
The three collision centralities include the centrality extremes (most
central and most peripheral) and an intermediate centrality that
requires a separate azimuth-quadrupole model element in the data
model.

Model functions include (a) a ``basic Model'' consisting of a
same-side (SS) peak modeled by a Gaussian at $\phi = 0$ and an
away-side (AS) peak at $\pi$ modeled by a cylindrical dipole
$\cos(\phi - \pi)$, (b) the basic Model plus one or more additional
cosine terms and (c) several Fourier-series (FS-only) models
consisting only of one or more cosine terms.

For each model-data combination we obtain the best-fit $\chi^2$ and
information $I$ and combine them to form {\em evidence} $E =
\exp[-(\chi^2/2 + I)]$ interpreted in a BI context as the probability
of data $D^*$ given model $H$. Information $I$ is the logarithm of a
volume ratio. The numerator is a ``prior'' volume on the space of
model parameters determined consistently from model to model based on
the nature of the parameters. The denominator is the volume on model
parameters determined by the fit covariance matrix. Thus, $I$ measures
information received by the model from new data and is interpreted in
the BI context as an {\em Occam penalty}, with reference to Occam's
razor. With increasing model complexity (degrees of freedom) $\chi^2$
typically decreases but $I$ increases, leading to a maximum in
evidence $E$ for some model configuration.

For each centrality we rank models according to evidence $E$ which can
vary over several orders of magnitude. The following systematics
emerge: FS-only models (c) are rejected in all cases by at least a
factor 100. The basic Model (a) representing peaks at 0 and $\pi$ is
preferred in all cases. A cylindrical quadrupole $\cos(2\phi)$ is
required to accompany the basic Model in some cases but is rejected
for most-central \auau\ collisions.  ``Higher harmonics''
$\cos(m\phi)$ for $m > 2$ appended to the basic Model are rejected in
all cases.  A model consisting of Gaussian + dipole $\cos(\phi)$ +
quadrupole $\cos(2\phi)$ provides good data descriptions in all
cases. Those results are generally consistent with a {\em power
  spectrum} analysis of data histograms in which signal and noise
components are identified.

A detailed study of the geometric structure of Bayesian analysis
reveals that given comparable fit quality ($\chi^2$) for various data
models the dominant factor in determining $E$ is the ratios of data
errors to parameter errors. Those ratios estimate elements of the
Jacobian matrix characterizing the model function as a map from
parameters to data. Smaller error ratios indicate smaller predicted
volumes in the data space: the data model is more predictive. {\em
  Predictivity} is then the determining factor in Bayesian model
evaluation. FS-only models have no predictivity, can describe any data
configuration and are strongly rejected by Bayesian analysis. The
basic Model is very predictive and is therefore strongly favored.

We conclude from this study that QGP/flow narratives based on FS-only
models or models with multiple cosine terms are disfavored because the
requisite data models are rejected by Bayesian analysis. FS-only
models are not predictive, in particular cannot exclude dijets as a
dominant collision mechanism. Dijet-based narratives are favored in
that the basic Model, with peaks at 0 and $\pi$ that may represent
dijet structure expected in such narratives, is strongly preferred by
Bayesian analysis.
This conclusion should of course be tested more generally with other
data and contexts such as unprojected 2D angular-correlation
histograms and their corresponding more-complex model
parametrizations.

\vskip .1in

%%%%%%%%%%%%%%%%%%
\textbf{Acknowledgments}: This work is supported in part by a
Consolidoc fellowship, the National Institute for Theoretical Physics
and the National Research Foundation of South Africa.

\begin{appendix}

%%%%%%%%%%%%
\section{BI analysis geometry} \label{structure}

Section II indicates that BI analysis provides two important results:
(a) an improved posterior PDF on model parameters given newly-acquired
data and (b) a quantitative method for comparing data models to
identify the model function that achieves the best compromise between
accurate data description and minimum Occam penalty. In this appendix
we examine the geometric structure of BI analysis on the joint
parameter-data space to better understand how the BI method works.
We find that evidence $E(D^*|H)$ is a measure of the
\textit{predictivity} of a model: BI analysis prefers the most
predictive model that also describes the data with a satisfactory
$\chi^2$.

\subsection{Data space vs model-parameter space}

BI analysis is based on the relation between parameter space $w$ and
data space $D$.  The data space is an N-dimensional space with axes
$D_n$.
The model-parameter space is a K-dimensional space with axes
$w_k$. Data model $H$ is defined in part by model function $F(D|wH)
\rightarrow D(w)$ that relates a specific set of parameter values
(point $w^*$ in $w$) to a specific set of data values (point $D^*$ in
$D$).
Note that data $D$ and data errors $\sigma_D$ are vectors with
elements $D_n$ and $\sigma_n$. Similarly, parameters $w$ and parameter
errors $\sigma_w$ are vectors with elements $w_k$ and $\sigma_k$.

As set out in Sec.~\ref{chain}, a joint parameter-data PDF $p(wD|H) =
p(D|wH)p(w|H)$ representing model $H$ can be defined on the joint
space $w\times D$.  If data values originate as random samples from
some parent distribution related to model $H$ with specific parameter
values $w^*$ the resulting data distribution may be described by a
localized conditional PDF $p(D|w^*H)$ on $D$ with estimated means
$\tilde D_n$ and standard deviations $\sigma_n$. Conversely, for a
specific set of data values $D^*$ the resulting parameter distribution
may be described by a localized conditional PDF $p(w|D^*H)$ on $w$
with estimated means $\tilde w_k$ and standard deviations $\sigma_k$.

Whereas conditional PDFs $p(D|w^* H)$ and $p(w|D^*H)$ may be localized
near their respective modes $\tilde D$ and $\tilde w$, marginal PDFs
$p(w|H)$ and $p(D|H)$ (dash-dotted lines in Fig.~\ref{bayessp}) may be
nearly uniform over the local intervals relevant to the peaked
functions.
In what follows we extend the BI methodology to obtain a {\em global}
geometric relation between parameter space and data space pursuant to
model comparisons. We refer to conditional PDFs $p(D|wH)$ and
$p(w|DH)$ as local and marginal PDFs $p(w|H)$ and $p(D|H)$ as global.

\subsection{Angle representation of model structure} \label{anglex}

For each data model $H$ the primary BI elements are model function
$D(w)$, prior PDF $p(w|H)$ (assuming a uniform prior on parameters
$w_k$), some specific data $D^*$ and their uncertainties $\sigma_D$. A
joint PDF $p(wD|H)$ determined by function $D(w)$ {\em and errors}
$\sigma_D$ is then implicit.
In a model fit to some specific data $D^*$ the data errors $\sigma_D$
and model function $D(w)$ are combined to determine the most-probable
model parameters $\tilde w$ and their uncertainties $\sigma_w$, or
preferably a posterior PDF $p(w|D^*H)$ on space $w$ (as in
Sec.~\ref{bayes1}).

%%%%%%%%%%%%%%%%%%%%%%%
\begin{figure}[h]
%\put(-28,90) {\bf (a)}
\includegraphics[width=.48\textwidth]{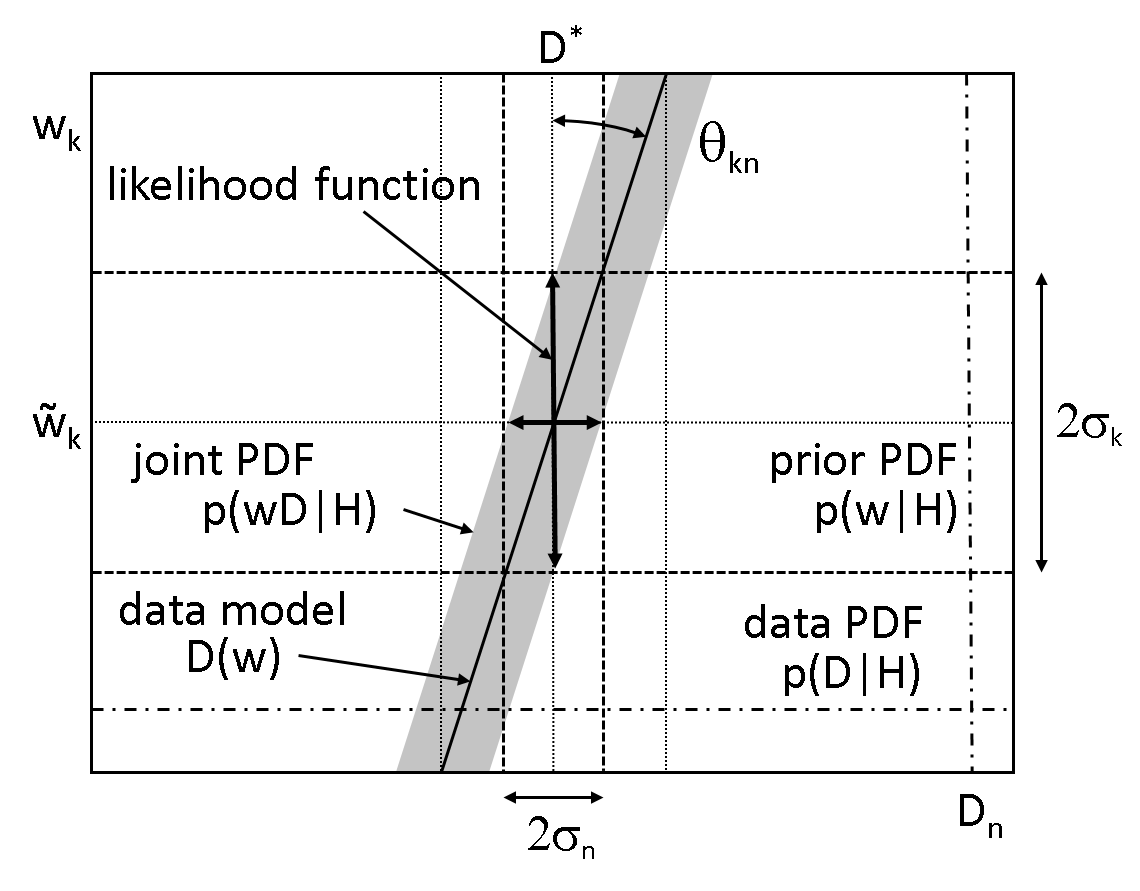}   
%\put(-158,120) {\bf bin 10}
\caption{\label{bayessp} Schematic representation of the local
  relation in space $w \times D$ between data $D$ and model parameters
  $w$ with specific elements $D_n$ and $w_k$, especially the
  errors. The solid diagonal represents model function $D(w)$. The
  hatched band arises from data errors, specifically $\sigma_n$,
  corresponding then to parameter errors, specifically
  $\sigma_k$. Angle $\theta_{kn}$ relating data and model errors is
  approximately a Jacobian element characterizing the algebraic
  structure of the data model. The dash-dotted lines represent
  parameter-prior and data PDFs.  } %bayes-space
\end{figure}
%%%%%%%%%%%%%%%%%%%%%%%

Figure~\ref{bayessp} provides a schematic of data-model
correspondence, with model parameter $w_k$ and data element $D_n$ in
the local neighborhood of specific data values $D^*$. The diagonal
line represents the model function $D(w)$. The data values $D_n^*$
have estimated standard deviations $\sigma_n$ (data errors). The model
function with data errors determines the gray band representing joint
PDF $p(wD|H)$.

The likelihood function $L(D^*|w H)$ on $w_k$ determines the
most-probable parameter values $\tilde w_k$ and their standard
deviations $\sigma_k$ corresponding to data values $D^*$ and data
errors $\sigma_D$.
As indicated by the bold vertical arrow in Fig.~\ref{bayessp} the
likelihood function, with specific data errors, in effect probes the
local algebraic structure of model function $D(w)$ near data $D^*$ by
relating data errors $\sigma_n$ to parameter errors $\sigma_k$.
The geometric relation between data and parameters is characterized by
angles $\theta_{kn}$ defined by
\begin{eqnarray}
  \tan(\theta_{kn}) &\equiv& \frac{\langle \sigma_n\rangle}{\sigma_k}
\end{eqnarray}
that relate data and parameter spaces. If the Hessian matrix for this
application is diagonal (i.e.\ correlations among model parameters are
small) those angles correspond to elements of the model-function
Jacobian $J_{D(w)}$
\begin{eqnarray} \label{angles} \tan(\theta_{kn}) \approx \sqrt{N}
  \left\langle \frac{\partial D_n}{\partial w_k}\right\rangle
  \leftrightarrow J_{D(w)},
\end{eqnarray}
in the following sense: If the diagonal elements of the Hessian are
approximated by
\begin{eqnarray}
  -H_{kk} = \frac{1}{\sigma_{k}^2}
  &\simeq& N \left \langle \frac{1}{\sigma^2_n} \right \rangle \left\langle
    \left( \frac{\partial D_{n}(w)}{\partial w_k} \right)^2 \right\rangle_{\!\!n}
\end{eqnarray}
the partial derivative in Eq.~(\ref{angles}) represents an r.m.s.\
quantity derived by averaging squared Jacobian elements over all data
elements (weighted by the data errors).  The same Jacobian structure
may determine the relation between global structures (PDFs) on $w$ and
on $D$ by extrapolation, as discussed in the next subsection.

\subsection{Local and global volumes vs PDFs}

We can define effective volumes (generalized concept including
lengths, areas, etc.) in spaces $w$ and $D$ in relation to the key
PDFs associated with BI. By volume we mean the result of integrating a
unit-amplitude (at the mode) function over some bounded subspace
including all points where the function is significantly nonzero. For
example the ``volume'' of unit-amplitude 1D Gaussian
$e^{-x^2/2\sigma^2}$ is $\sqrt{2\pi}\sigma$. Dividing a unit-amplitude
function by its volume results in a normalized PDF.

The marginal PDF on data $D$ is obtained by integrating $p(wD|H)$ over
space $w$ using the chain rule
\begin{eqnarray} \label{pdh} 
  p(D|H) &=& \int dw\, p(D|wH)\, p(w|H) \\
  \nonumber &=& \frac{1}{V_w(H)} \int_{V_w} dw\, p(D|wH).
\end{eqnarray}
The (assumed uniform) prior PDF $p(w|H)$ defines an effective boundary
surface for the integral over $w$, represented schematically by the
second line where each parameter $w_k$ is integrated over a prior
interval $\Delta_k$ and $ \prod_{k=1}^K \Delta_k \equiv V_w(H) =
1/p(w|H)$ as in Eq.~(\ref{priordelt}).

For some values of $D$ the integrand may be nonzero only outside the
volume $V_w(H)$, in which case $p(D|H) = 0$. If $p(D|H)$ is nonzero
and approximately uniform within limiting intervals $\Delta_n$ then
$p(D|H) \approx 1/V_D(H)$ with $V_D(H) = \prod_{n=1}^N \Delta_n$, and
Eq.~(\ref{pdh}) represents a relation between the two global volumes
$V_w(H)$ and $V_D(H)$.

The dual role of $p(D|wH)$ as conditional PDF on $D$ and as likelihood
function on $w$ is a central issue. With $w^*$ as a specific condition
$p(D|w^*H)$ is a {\em unit-normal} peaked distribution on $D$
approximated by a Gaussian
\begin{eqnarray} \label{pdw}
  p(D|w^* H) &\approx& \frac{1}{V_D(w^*|H)} G(D|w^*),
\end{eqnarray}
where $V_D(w^*|H) \equiv \prod_{n=1}^N [\sqrt{2\pi} \sigma_n]$.
As the likelihood function $L(D^*|w H)$ it is an unnormalized peaked
distribution on parameter space $w$ which in the Laplace approximation
is proportional to Gaussian $G(w|D^*)$ with its integral
\begin{eqnarray}
\int dw \, G(w|D^*) &\approx& V_w(D^*|H) \equiv \prod_{k=1}^K [\sqrt{2\pi} \sigma_k],
\end{eqnarray}
where $V_w(D^*|H)$ approximates $ \sqrt{(2\pi)^K \det C_K}$ appearing
in Eq.~(\ref{evid}).  We have thus defined four volumes, two each on w
and D: two local and two global.

\subsection{Consequences for the data space -- predictivity}

Using the Laplace approximation the BI evidence as defined in Eq. (5)
can be written in terms of volumes as %
\begin{eqnarray}
  E(D^*|H) &=&  \int dw\, L(D^*|wH)\, p(w|H)
  \\ \nonumber
  &\approx&  L(D^*|\tilde w H)  \frac{V_w(D^*|H)}{V_w(H)}.
\end{eqnarray}
From Eq.~(\ref{pdw}) the maximum likelihood $L(D^*|\tilde w H)$ is
\begin{eqnarray}
  p(D^*|\tilde w H) &\approx& \frac{G(D^*|\tilde w)}{V_D(\tilde w|H)}.
\end{eqnarray}
with $G(D^*|\tilde w) = \exp(-\chi^2/2)$.
If $D^*$ falls outside $V_D(H)$ evidence $E(D^*|H) = 0$ (the data
model is falsified). If not $E(D^*|H) \approx p(D|H) \approx 1/V_D(H)$
and we then have
\begin{eqnarray} \label{edh}
  \frac{V_D(\tilde w|H)}{V_D(H)} &\approx&G(D^*|\tilde w)  \frac{V_w(D^*|H)}{V_w(H)}
  \\ \nonumber 
  &\approx& \exp[-(\chi^2/2+I)],
\end{eqnarray}
(with information $I$ as defined in Sec.~\ref{modcompare}) relating
the four volumes, where factor $V_D(\tilde w|H)$ is a property of the
data only, common to all models.

We can relate that result to the model angles (Jacobian) from
Sec.~\ref{anglex}. Assuming data errors $\sigma_n$ are approximately
equal the local-volume ratio is factorized as
\begin{eqnarray}
  \frac{V_w(D^*|H)}{V_D(\tilde w|H)} &=& \frac{ \prod_{k=1}^K [\sqrt{2\pi}\sigma_k]}{\prod_{n=1}^N [\sqrt{2\pi}\sigma_n]} 
  \\ \nonumber
  &\approx& \frac{1}{V_D(\tilde w|H)^{(N-K)/N}} \prod_{k=1}^K\frac{1}{\tan(\theta_{kn})},
\end{eqnarray}
where the first factor in the second line depends only on data\ common
to all models with parameter number $K$, and the second factor is
unique to a specific model.

Rearranging Eq.~(\ref{edh}) (without Gaussian factor) as
\begin{eqnarray} \label{vwh}
\frac{V_w(H)}{V_D(H)} &\approx&\frac{V_w(D^*|H)}{V_D(\tilde w|H)}
\end{eqnarray}
we note that the local-volume ratio on the right, obtained from the
likelihood function and equivalent to the model-function Jacobian,
estimates the global-volume ratio on the left by extrapolation.
$V_D(H)$ is then the data volume {\em predicted} by a combination of
prior PDF on model parameters and the model function.
If the specific data values $D^*$ fall outside $V_D(H)$ then $p(D|H) =
0$ and the model is falsified. The smaller the predicted data volume
the larger the evidence and the more favored the model. {\em
  Predictivity} is then any essential feature of data models.

The Occam penalty central to BI analysis represents not only excess
parameter number and prior volume as a cost but also model
predictivity as a benefit. Two models with the same parameter number
and priors may have very different plausibilities because of
differences in their algebraic structure and therefore predictivity. A
model with substantially greater predictivity may even be favored over
one with fewer parameters (Sec.~\ref{bin0}).

%%%%%%%%%%%%
\section{Hessian matrix errors} \label{hesserror}

Since comparisons among data models in this BI study rely critically
on fitted-parameter errors it is important to establish the degree of
uncertainty in the estimated errors.  Parameter variances are
determined by the curvatures (second derivatives) of the log-likehood
function at the likelihood maximum. The Hessian is defined by
\begin{eqnarray}
  H_{jk} = \left. \frac{\partial^2}{\partial w_j\partial w_k}\log L(D^*|wH)\right|_{w=\tilde{w}}.
\end{eqnarray}
For linear parameters such as the coefficients of a Fourier series the
second derivatives are independent of the parameter values
$w$. Specifically, for
\begin{eqnarray}
  &&	\hspace{-1.65in}	\log L(D^*|wH) \approx -\sum_{n=1}^N \frac{1}{2}\left(\frac{y_n - \sum_k w_k f_k(x_n)}{\sigma_n}\right)^2  \\
  \frac{\partial^2}{\partial w_j\partial w_k} \log L(D^*|wH) &\approx& -\sum_{n=1}^N \frac{f_j(x_n)f_k(x_n)}{\sigma_n^2}.
\end{eqnarray}
If the model functions are orthogonal we have
\begin{eqnarray}
  \sum_{n=1}^N \frac{f_j(x_n)f_k(x_n)}{\sigma_n^2} \rightarrow \delta_{jk} \sum_{n=1}^N \frac{f_j^2(x_n)}{\sigma_n^2}
\end{eqnarray}
and the Hessian matrix is diagonal.  In other words, for linear
parameters such as the coefficients of a Fourier series the parameter
variances depend only on the sample points (positions) and the
algebraic structure of the model function. For a Fourier series there
is little flexibility -- since there is no uncertainty in the model
function the only uncertainty in the Hessian arises from the
uncertainty in the sample positions.
Labeling the uncertainties in the $x$ coordinate due to bin width
$\delta x$ as $\sigma_{x_n} = (\delta x)^2/12$, the uncertainty in the
diagonal Hessian elements is
\begin{eqnarray}
  \sigma_{H_{jj}}^2 = \sum_{n=1}^N \left[\frac{2 f_j(x_n) f_j'(x_n)}{\sigma_n^2}\right]^2\frac{(\delta x)^2}{12}.
\end{eqnarray}
There is only one nonlinear parameter, namely the width $\sigma_\phi$
of the SS Gaussian in the basic Model, but a similar formula should
apply to that case as well.

For bin 8 the diagonals of Hessian and errors for the FS-only model are
\begin{eqnarray}
  H_{jj} &=& \{8.19, 8.22, 8.23, 8.23, 8.23,8.23, \\
  && 8.23, 8.23, 8.23, 8.22,8.19\} \times 10^6 \\
  \sigma_{H_{jj}} &=& \{3.46,6.91,10.4,13.8,17.3,0.00145,\\
  && 24.2,27.7,31.1,34.6,38.\} \times 10^4.
\end{eqnarray}
The relative errors thus vary from 0.5 percent up to five percent. For
the basic-Model fit in bin 8 we obtain
\begin{eqnarray}
  \bm{w} &=& \{a_1 , \sigma_{\phi},a_d\} \\
  H_{w} &=& \{0.478,1.55,8.19\} \times 10^6 \\
  \sigma_{H_{w}} &=& \{1.05,3.1,3.47\} \times 10^5,
\end{eqnarray}
implying relative errors from 5 percent to 22 percent. Thus the
Hessian matrix elements (likelihood curvatures), and therefore the
error estimates for fitted model parameters, are determined to a few
percent in this study.

%%%%%%%%%%%%
\section{Laplace approximation} \label{laplaceacc}

Use of the Laplace approximation for the likelihood function in this
study may be questioned due to possible inaccuracies.  The Laplace
approximation for the likelihood function $\log L(D^*|wH) \equiv
-g[{w}]$ is
\begin{eqnarray}
  g[{w}] &\approx& \sum_{n=1}^N \frac{1}{2}\left(\frac{y_n - \sum_k f_k(\bm{w},x_n)}{\sigma_n}\right)^2 \\
  && \hspace{-.5in}	= \hspace{.05in}  g[\tilde{w}] + (w-\tilde{w})\frac{\partial g[\tilde{w}]}{\partial w} 
  + \frac{(w-\tilde{w})^2}{2}\frac{\partial^2g[\tilde{w}]}{\partial w^2} \\ \nonumber
  &&  \hspace{-.5in} + \hspace{.05in} \frac{(w-\tilde{w})^3}{6}\frac{\partial^3g[\tilde{w}]}{\partial w^3}  + \cdots,
\end{eqnarray}
where the first-derivative term in the Taylor series is zero by
definition. The approximation then implies
\begin{eqnarray}
  \int_a^b dw\, e^{-g[w]} 
  &\approx& \int_{-\infty}^\infty dw\, e^{-g[\tilde{w}]-\frac{(w-\tilde{w})^2}{2}\frac{\partial^2g[\tilde{w}]}{\partial w^2}} \\ \nonumber
  &\times&	\left(1-\frac{(w-\tilde{w})^4}{24}\frac{\partial^4 g[\tilde{w}]}{\partial w^4} + \cdots \right) \\ \nonumber
  && \hspace{-.2in} \approx \hspace{.05in} e^{-g[\tilde{w}]}\sqrt{\frac{2\pi}{\frac{\partial^2g[\tilde{w}]}{\partial w^2}}}
  \left(1 - \frac{\frac{\partial^4 g[\tilde{w}]}{\partial w^4}}{8\left(\frac{\partial^2g[\tilde{w}]}{\partial w^2}\right)^2}\right),
\end{eqnarray}
where we have carried the first correction term. For linear
parameters, including all of the parameters except $\sigma_\phi$,
there is no fourth derivative as we have pointed out. The Laplace
approximation is then exact (except for sub-exponential corrections
caused by replacing the truncated Gaussian by a non-truncated
version). For the single non-linear parameter in the basic Model the
fourth derivative $1.09\times 10^6$ must be divided by $1.55\times
10^6$ squared, implying that corrections are of the order $10^{-5}$ --
of the same order as the fit-model Hessian. The effective number of
data points is $10^5$ and any corrections are of inverse order. Due to
a large primary-data volume [more than a million \auau\ collisions
with (on average) hundreds of particles per collision] the Laplace
method as applied in the present study is very accurate.

%%%%%%%%%%%%
\section{Periodic peak arrays} \label{periodpeak}

Because azimuth $\phi$ is a periodic variable any 1D structure on
\dphi\ can be described by a discrete Fourier cosine series FS.  But
representing an arbitrary 1D projection by a few terms of a 1D Fourier
series can be misleading. We should acknowledge the possibility that
specific peak structures may be part of the azimuth distribution.
In this appendix we consider the FS representation of a periodic
Gaussian peak array on 1D azimuth.

The peaks observed at $\phi_\Delta = 0$ (SS, same-side) and
$\phi_\Delta = \pi$ (AS, away-side) in all 1D azimuth histograms from
high-energy nuclear collisions are actually elements of separate
periodic peak arrays described by cosine {series}. The SS array is
centered on {even} multiples of $\pi$, the AS array on {odd}
multiples. Nearest array elements outside a $2\pi$ interval (image
peaks) produce significant structure within the observed interval and
must be included in fit models to insure valid data descriptions.

Each peak array (SS or AS) may be represented by a FS of the form
\begin{eqnarray} \label{fourier1}
S(\phi_\Delta;\sigma_{\phi_\Delta},n) \hspace{-.06in} &=& \hspace{-.09in}  \sum_{m=-N/2}^{N/2} \hspace{-.05in} F_{m,n}\, \cos(m\,[\phi_\Delta - n\,\pi]),
\end{eqnarray}
where the $F_{m,n}$ are functions of r.m.s.\ peak width
$\sigma_{\phi_\Delta}$ defined below.
Since $n$ is even for SS peak arrays ($+$) and odd for AS arrays ($-$)
odd multipoles must be explicitly labeled as SS or AS. The terms
represent $2m$ poles, e.g. dipole ($m=1$), quadrupole ($m=2$),
sextupole ($m=3$) and octupole ($m = 4$), referring to cylindrical
multipoles.

%%%%%%%%%%
 \begin{figure}[h]
  \includegraphics[width=1.65in,height=1.65in]{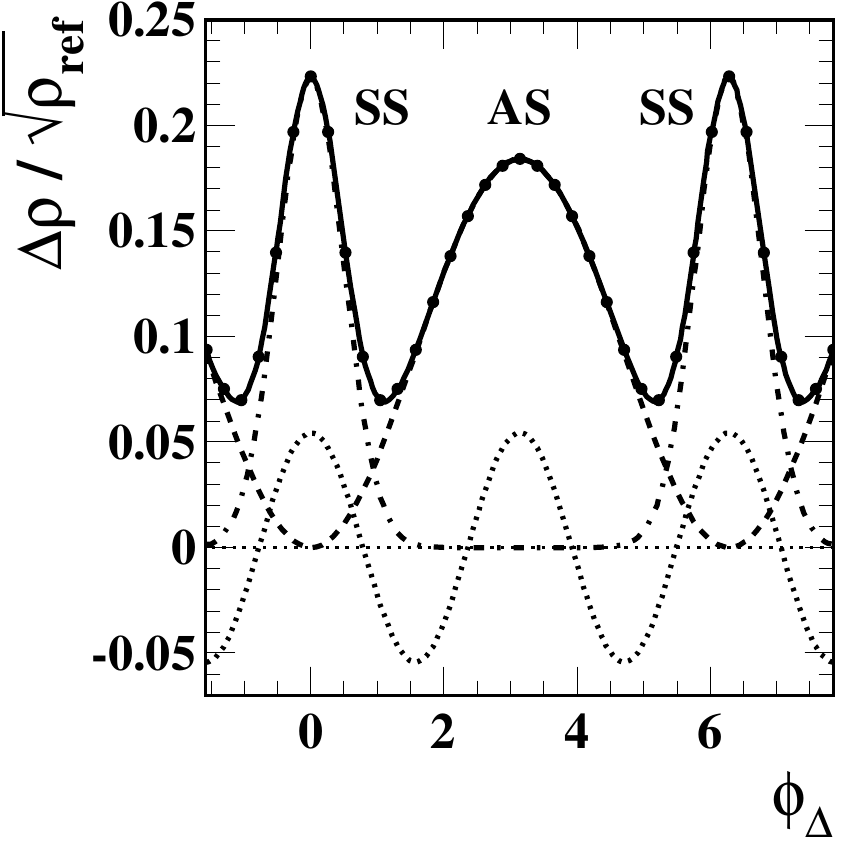}
  \includegraphics[width=1.65in,height=1.62in]{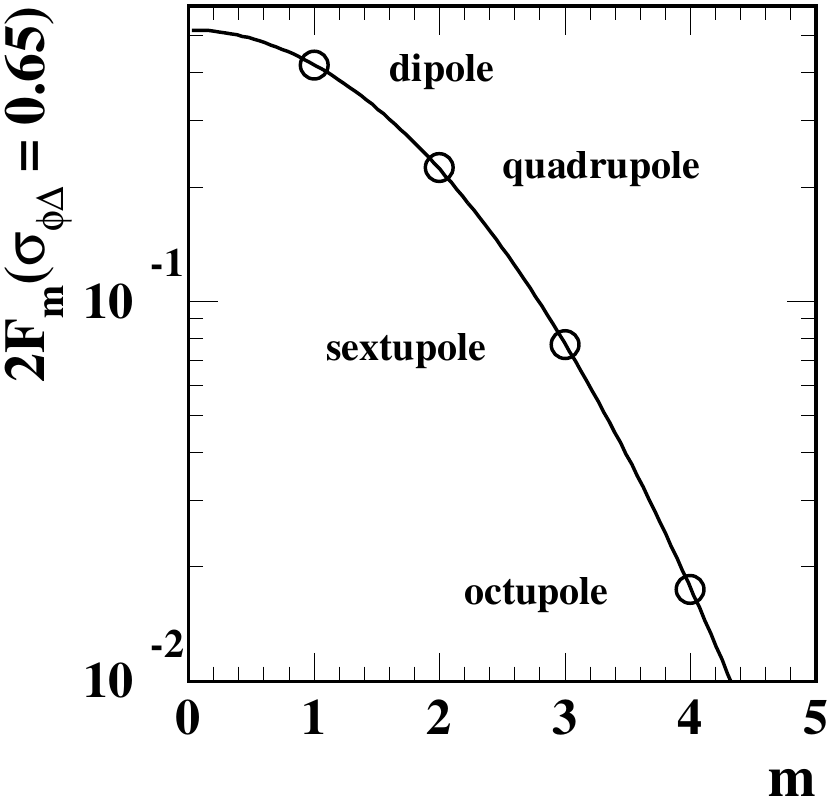}
  \caption{\label{ortho} Left: Periodic arrays of SS (dash-dotted) and
    AS (dashed) peaks. The SS peaks are Gaussians. The AS peaks are
    described by a dipole. The dotted sinusoid corresponds to the
    $m=2$ Fourier component of the SS peaks.  Right: Evaluation of
    Eq.~(\ref{fm}) for four values of $m$, with $\sigma_{\phi_\Delta}
    = 0.65$.  } %  newflow30aes, newflow35bx
 \end{figure}
%%%%%%%%%%%%

 The Fourier amplitudes $F_m$ of a unit-amplitude Gaussian peak array
 are defined (for $m \neq 0$) as functions of the r.m.s.~peak width by
\begin{eqnarray} \label{fm}
  2F_m(\sigma_{\phi_\Delta}) &=& \sqrt{2/\pi}  \sigma_{\phi_\Delta} \exp\left( - m^2 \sigma_{\phi_\Delta}^2 / 2\right).
\end{eqnarray}
As peak width $\sigma_{\phi_\Delta}$ increases, the width on index $m$
decreases and the number of significant terms in the series
Eq.~(\ref{fourier1}) decreases. The limiting case is
$\sigma_{\phi_\Delta} \approx \pi / 2$, for which the peak array is
approximated by a constant plus dipole term.
For narrower (SS) peaks Fourier terms with $m > 1$ become significant,
and a Gaussian function is the more efficient peak model, as
demonstrated in this study.

Fig.~\ref{ortho} (left panel) shows peak arrays (solid points) for SS
and AS peaks extending beyond one $2\pi$ period. The SS Gaussian peak
array with $\sigma_{\phi_\Delta} = 0.65$ (typical value for all but
peripheral \aa\ collisions) is the dash-dotted curve, the AS array
with $\sigma_{\phi_\Delta} \sim \pi / 2$ is the dashed curve
(approximately dipole in this case). The dotted curve represents the
quadrupole term of the SS peak array.

Figure~\ref{ortho} (right panel) shows Eq.~(\ref{fm}) for
$\sigma_{\phi_\Delta} = 0.65$ with the first few multipole
coefficients marked for reference (open circles).
For that width the jet-related quadrupole amplitude is $2F_2 \approx
0.22$. If the SS peak is not separately described by a Gaussian peak
model $F_2$ represents the dominant jet-related nonflow contribution
to $v_2^2\{2\} \sim v_2^2\{EP\}$ data in the form $\rho_0
v_2^2$. Similarly, other Fourier components of the SS jet peak and the
dipole component of the AS peak could be misidentified as flow
components, including ``higher harmonic'' flows~\cite{multipoles}.

\end{appendix}

%%%%%%%%%%%%%%%%%%%%%%%%%%%%%%%%%%%%%%%%%%%%%%%%%%%%%%%%%%%%%%%%%%%%%%%%%%%

\end{document}